\documentclass{aa}  

\usepackage{graphicx}
\usepackage{txfonts}
\usepackage[colorlinks=true, citecolor=blue, urlcolor=blue]{hyperref}
\usepackage{placeins}
\usepackage{color, soul}
\usepackage{ulem}
\usepackage{natbib}
\usepackage{lipsum}
\bibpunct{(}{)}{;}{a}{}{,} % to follow the A&A style
\usepackage{tabularx}  % Add this to your preamble if not already included
\usepackage{caption}
\usepackage[switch]{lineno}

\begin{document} 

%\linenumbers

   \title{Investigating the emission signatures of pulsar halo candidate HESS J1813-126}

     \titlerunning{Investigating pulsar halo candidate HESS J1813-126}

   \author{Agnibha De Sarkar
          }

   \institute{Institute of Space Sciences (ICE, CSIC), Campus UAB, Carrer de Can Magrans s/n, E-08193 Barcelona, Spain \\
             \email{desarkar@ice.csic.es}\\
            }
    
   \date{Received XXXX; accepted YYYY}
% \abstract{}{}{}{}{} 
% 5 {} token are mandatory
 
\abstract
   {Extended gamma-ray sources surrounding middle-aged pulsars, primarily observed at teraelectronvolt energies, have been interpreted as pulsar halos, where relativistic $e^\pm$ diffuse into the interstellar medium and produce inverse-Compton (IC) emission. HESS J1813-126, associated with the energetic, radio-quiet gamma-ray pulsar PSR J1813-1246, has been suggested as a candidate pulsar halo, though its nature remains uncertain.}
   {We interpreted the high-energy emission of PSR J1813-1246 using the synchro-curvature (SC) radiation model and tested whether the gamma-ray spectral energy distribution (SED) of HESS J1813-126 can be explained as a pulsar halo powered by PSR J1813-1246.}
   {We explain the X-ray and gamma-ray SEDs of the pulsar using the SC framework. We further computed the transport and losses of $e^\pm$ injected by the pulsar through time-dependent diffusion-loss equations, exploring various common pulsar halo transport models. The resulting IC emission was compared with \textit{Fermi}-LAT, H.E.S.S., HAWC, and LHAASO data. We present predictions for the surface brightness profiles (SBPs) and the aperture-dependent emission for the different transport models, providing key diagnostics for assessing the observability of HESS J1813-126 with current and future instruments.}
   {The SC framework successfully reproduces the emission of PSR J1813-1246. The SED of HESS J1813-126 can be consistently reproduced within different pulsar halo frameworks, albeit with distinct predictions across different transport models. The corresponding SBP predictions and aperture-dependent emission offer testable signatures for future imaging atmospheric Cherenkov telescopes, which will be crucial for discriminating between the transport models. We further examined the link between the pulsar central engine and its extended halo by comparing the pair multiplicities in the magnetospheric and halo regions. 
   }
   {}


   \maketitle
%
%-------------------------------------------------------------------

\section{Introduction}

Extended gamma-ray emission around pulsars has emerged as a phenomenon distinct from compact pulsar wind nebulae (PWNe), which are typically confined within a few parsecs \citep{gaensler06,torres14}. These extended sources, termed pulsar halos (or sometimes teraelectronvolt halos), span degree scales without clear boundaries; they were first revealed by the High Altitude Water Cherenkov (HAWC) around Geminga and Monogem \citep{abey17,albert24}, and were interpreted as relativistic $e^\pm$ diffusing into the interstellar medium (ISM) and producing gamma-ray emission via inverse-Compton (IC) scattering of the background radiation field \citep{sudoh19}. The bright, extended teraelectronvolt (TeV)  morphologies of these halos (tens of parsecs) indicate suppressed diffusion coefficients \citep{abey17} relative to the Galactic average extrapolated from the boron-to-carbon (B/C) ratio \citep{genolini19,desarkar21}. However, alternative explanations are available (see, e.g., \citealt{liu19, aloisio09}, with detailed reviews in \citealt{fang22, liu22}).  

These halos may help explain the long-standing cosmic-ray positron excess problem observed by Payload for Antimatter Matter Exploration and Light-nuclei Astrophysics (PAMELA), Alpha Magnetic Spectrometer (AMS-02), and \textit{Fermi} Large Area Telescope (LAT) \citep{adriani09,aguilar13,ackermann12}, as nearby sources like Geminga and Monogem could dominate the local $e^\pm$ flux \citep{hooper17,manconi20}, though dark matter annihilation, and secondaries produced in supernova remnants (SNRs) or molecular clouds (MCs) remain viable scenarios \citep{hooper04,cholis09,cholis13,mertsch14,desarkar21}. The presence of halos also suggests that many unidentified extended TeV sources are halos rather than shell-type SNRs or PWNe \citep{giacinti20,martin22a,martin22b, lopezcoto18}, as halos are potentially common around middle-aged ($> 20 \ \mathrm{kyr}$) pulsars \citep{albert25}. It is indeed a possibility that new halo candidates will continue to emerge in \textit{Fermi}-LAT \citep{dimauro19,dimauro21}, High Energy Stereoscopic System (H.E.S.S.) \citep{hess18}, HAWC \citep{3hwc20}, and Large High Altitude Air Shower Observatory (LHAASO) catalogs \citep{1lhaaso24}. LHAASO detections above 100~TeV further highlight the prevalence of Galactic ``PeVatrons,'' traditionally attributed to SNR-MC systems or PWNe (see, e.g., \citealt{kar22,desarkar22a,desarkar22b,desarkar23,desarkar24, joshi23, wu23, mitchell24, mitchell24b, abe23, delafuente23}), but pulsar halos themselves may also qualify if source confusion in the crowded inner Galaxy can be resolved \citep{mestre22}.  

Within this context, PSR J1813-1246 and its associated TeV source HESS J1813-126 offer a clean case study due to their off-plane location ($b \approx 2.5^{\circ}$). PSR J1813-1246 was discovered with \textit{Fermi}-LAT \citep{abdo09} and is the fastest-spinning, second-most energetic radio-quiet pulsar, with period $P \simeq 48.1$ ms, period derivative $\dot{P} \simeq 1.76 \times 10^{-14}$ s s$^{-1}$, current spin-down energy-loss rate $\dot{E} \simeq 6.24 \times 10^{36}$ erg s$^{-1}$, and characteristic age $\tau_c \sim 43$ kyr \citep{abdo13}. The distance to the pulsar was estimated to be between a minimum of 2.5 kpc and a maximum of 24.7 kpc \citep{abdo13,marelli14}. Despite its energetics, no radio pulsations have been detected \citep{ray11, abdo13}. In X-rays, \textit{XMM-Newton }and \textit{Chandra} revealed a pulsed counterpart \citep{marelli14}. A possible detection of the pulsar in the 30-500 keV range was also reported using the International Gamma-Ray Astrophysics Laboratory (INTEGRAL) data, which is consistent with the extrapolation between the soft X-ray and gamma-ray spectra \citep{marelli14}.

The pulsar is positionally coincident with HESS J1813-126, a hard-spectrum ($E^{-2}$), extended ($\sim 0.21^{\circ}$) TeV source from the H.E.S.S. Galactic Plane Survey (HGPS; \citealt{hess18}). No gigaelectronvolt (GeV) or X-ray PWN associated with the pulsar has been detected \citep{ackermann11,marelli14,guevel25}, thus supporting a halo origin. HAWC detected the source as 3HWC J1813-125 \citep{3hwc20}, while LHAASO detected $>$100 TeV emission from the region as 1LHAASO J1813-1245 \citep{1lhaaso24}, confirming its extended morphology ($\lesssim 0.32^{\circ}$) and PeVatron potential. Together, these data establish HESS J1813-126 as a bright, spatially extended TeV gamma-ray emitter extending into the ultra-high-energy regime and naturally linked to PSR J1813-1246, thus making it a possible pulsar halo candidate.

In this paper we study the combined PSR J1813-1246 -- HESS J1813-126 system. Section \ref{results} presents and discusses the results of this work, and Sect. \ref{conclusion} concludes. Appendix \ref{lat} presents the details of the \textit{Fermi}-LAT standard and phased analyses of the GeV gamma-ray data from the pulsar and the search for a potential off-pulse emission. Appendix \ref{pulsar} discusses the synchro-curvature (SC) framework used to explain the high-energy emission of the pulsar, and Appendix \ref{halo} provides the formalisms of pulsar halo particle transport under the two-zone isotropic suppressed diffusion, ballistic-to-diffusion, and anisotropic diffusion scenarios.

\section{Results and discussion}\label{results}

\begin{figure*}
    \centering
    \includegraphics[width=0.33\textwidth]{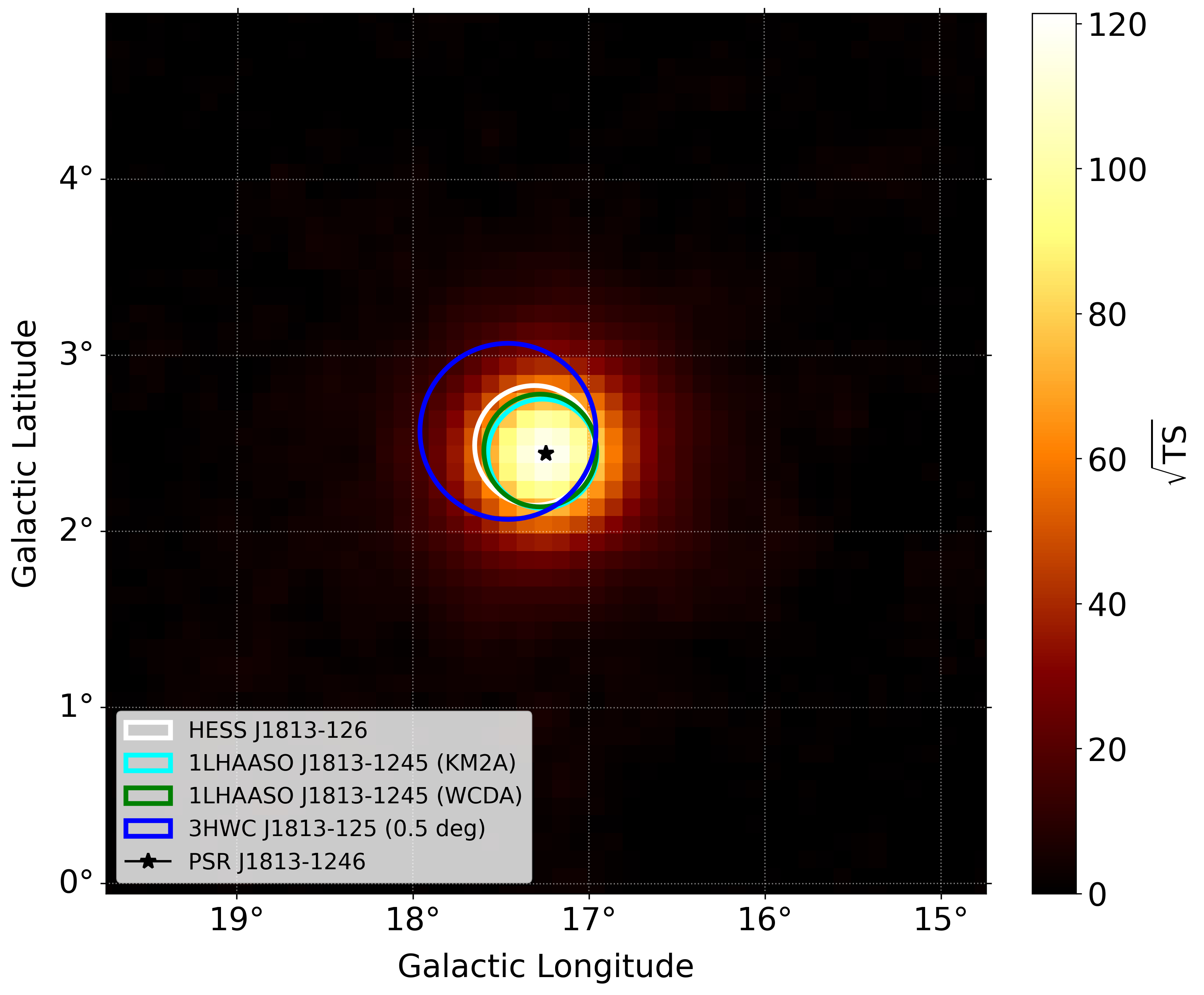}
    \includegraphics[width=0.33\textwidth]{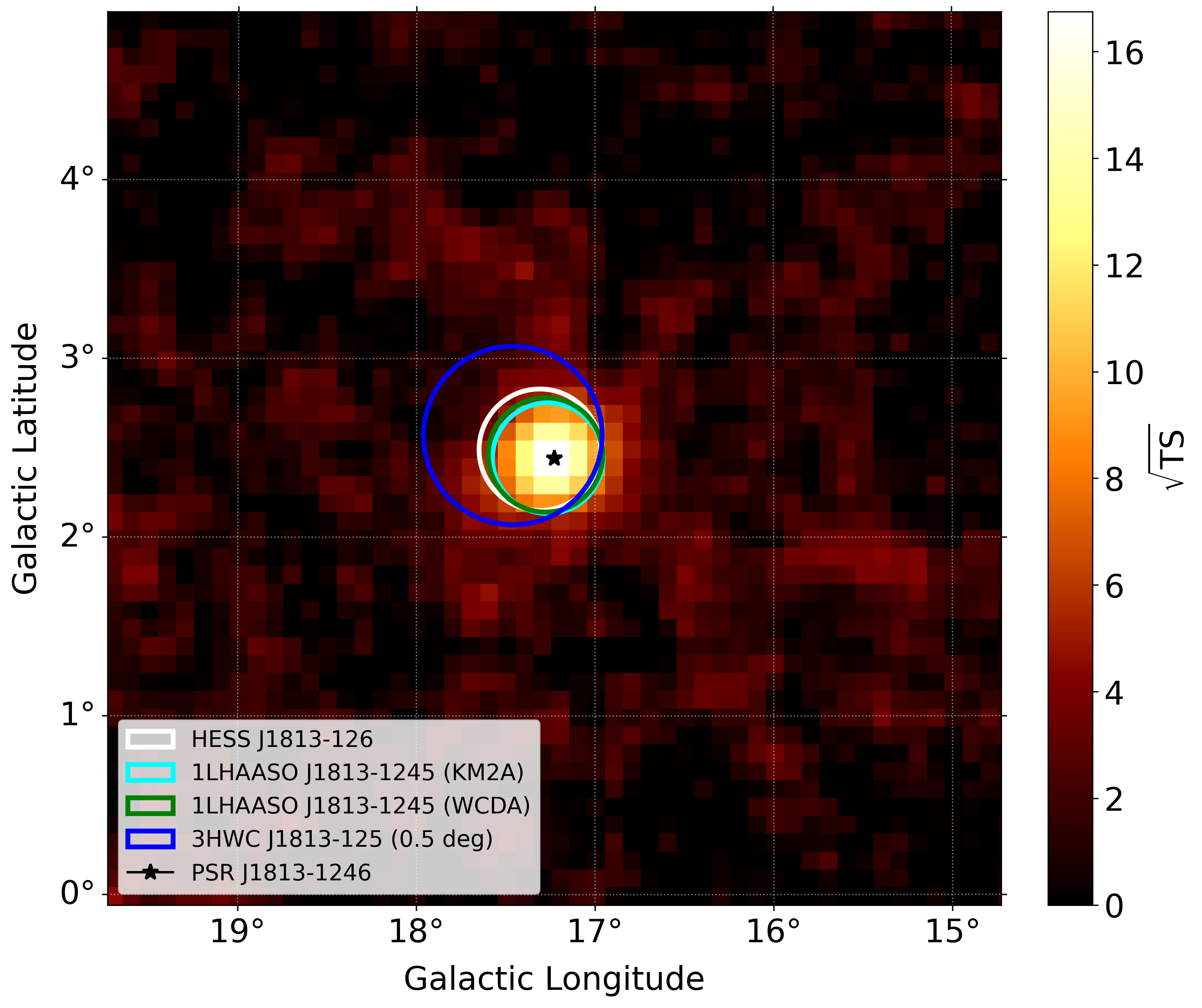}
    \includegraphics[width=0.33\textwidth]{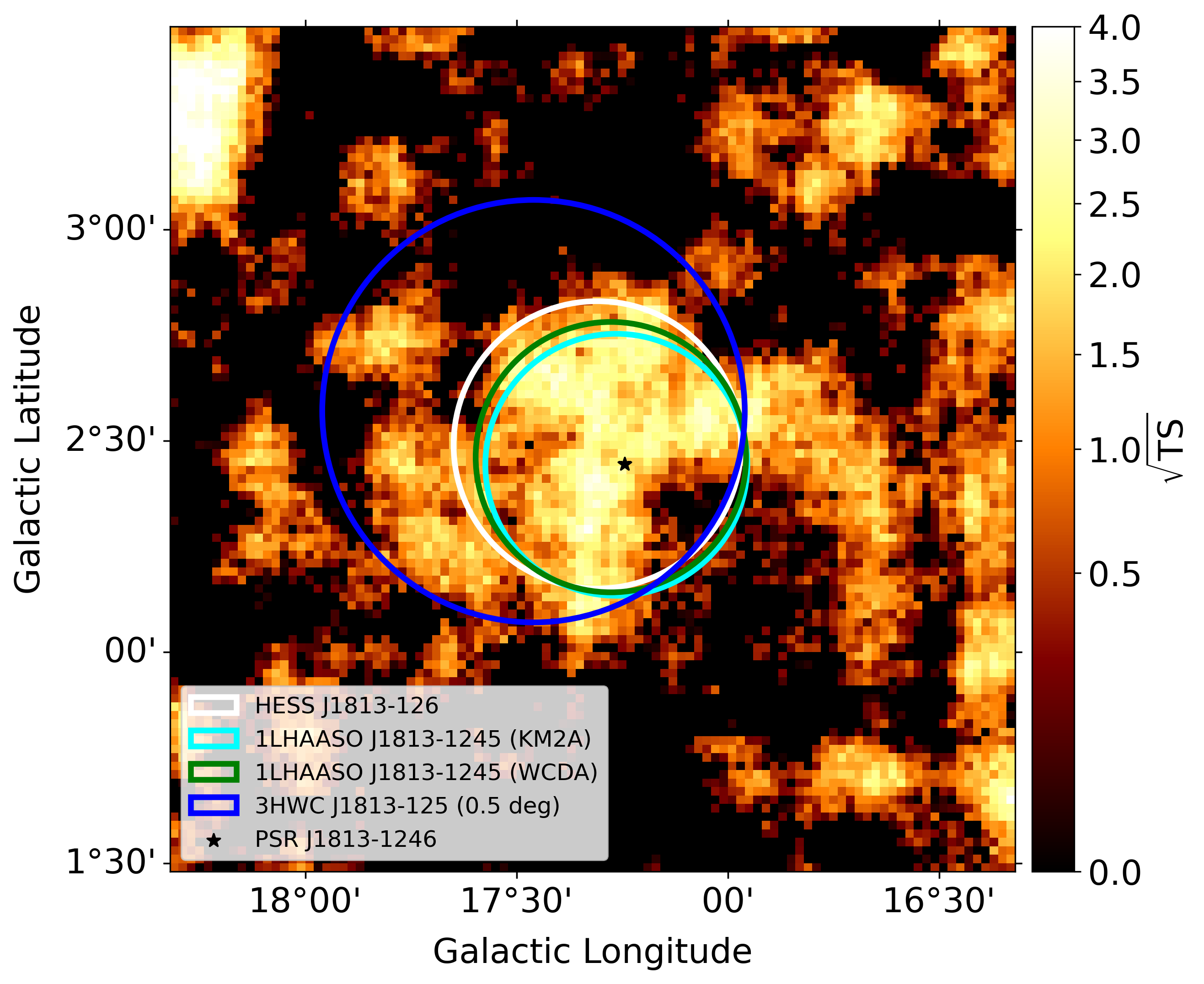}
    \caption{Significance maps of the HESS J1813-126 region. Left: Significance map of 4FGL J1813.4-1246 from the phase-averaged analysis. Middle: Same but for PS J1813.3-1246, detected during the off-pulse phase of the 4FGL source. Right: H.E.S.S. significance map. The color bar in each plot provides the significance level. The source extent for H.E.S.S., HAWC, and LHAASO (KM2A and WCDA) is shown as colored circles.}
    \label{fig: morpho}
\end{figure*}

The method for the standard \textit{Fermi}-LAT data analysis is discussed in Appendix \ref{lat}. PSR J1813-1246 is detected as 4FGL J1813.4-1246 in the updated Fourth \textit{Fermi} Gamma-ray LAT Source Catalog – Data Release 4 (4FGL-DR4). {The best-fit localization of the GeV counterpart was found at $\mathrm{RA} = 273.3471^{\circ} \pm 0.0025^{\circ}$ and $\mathrm{Dec} = -12.7667^{\circ} \pm 0.0025^{\circ}$, with a detection significance of $\sim 136\sigma$}. The best-fit position of the GeV counterpart is offset from the position of the pulsar by $0.08^{\circ}$, and is consistent with the centroid of the HESS J1813-126 ($\mathrm{RA} = 273.340^{\circ}$ and $\mathrm{Dec} = -12.688^{\circ}$), as well as being situated well within the containment radii of the HAWC and LHAASO detected counterparts.
The spectral analysis of the 4FGL source reveals a significantly curved spectral energy distribution (SED). The spectrum is best described by a \texttt{PLSuperExpCutoff4} function\footnote{\url{https://fermi.gsfc.nasa.gov/ssc/data/analysis/scitools/source_models.html}}, {which is 
favored over a simple power law function by $2 \log(\mathcal{L}_{\mathrm{PLSEC4}}/\mathcal{L}_{\mathrm{PL}}) \approx 1140.2$, and log-parabola function by $2 \log(\mathcal{L}_{\mathrm{PLSEC4}}/\mathcal{L}_{\mathrm{LP}}) \approx 982.8$.} This function is
particularly relevant for modeling the gamma-ray emission of pulsars. {The best-fit parameters of the \texttt{PLSuperExpCutoff4} function for this source are $\gamma_{\rm 0, \ \mathrm{PLSEC4}} = -2.459 \pm 0.001$, $E_{\rm 0, \ \mathrm{PLSEC4}} = 1440 \ \mathrm{MeV}$, $d_{\mathrm{PLSEC4}} = 0.667 \pm 0.001$ and $b_{\mathrm{PLSEC4}} = 0.426$}, where $\gamma_{0, \ \mathrm{PLSEC4}}$ and $d_{\mathrm{PLSEC4}}$ are the spectral index and local curvature at reference energy $E_{0, \ \mathrm{PLSEC4}}$. The index $b_{\mathrm{PLSEC4}}$ and reference energy $E_{0, \ \mathrm{PLSEC4}}$ were kept fixed at the 4FGL-DR4 catalog values. {The integrated energy flux between 1 and 500~GeV is $(8.304 \pm 0.095)\times10^{-11}~\mathrm{erg~cm^{-2}~s^{-1}}$.}  
We further examined the source extension (68$\%$ containment radius) using both \texttt{RadialDisk}
and \texttt{RadialGaussian} 
spatial models. {We found 
that \texttt{RadialDisk} model gives 
a best-fit extension of $0.0515^{+0.0098}_{-0.0109}$ deg with $\mathrm{TS_{ext}} \approx 7.169 \ (\sim 2.7 \sigma)$, 
where for \texttt{RadialGaussian}, these values are $0.0474^{+0.0092}_{-0.0099}$ deg with $\mathrm{TS_{ext}} \approx 7.307 \ (\sim 2.7 \sigma)$.} 
The source spatial model clearly agrees with the point-like hypotheses and disfavors the GeV source being extended. The curved SED and point-like morphology therefore confirm that 4FGL J1813.4-1246 is indeed the phase-averaged GeV counterpart of PSR J1813-1246.
Figure \ref{fig: morpho} shows the significance map of the phase-averaged 4FGL J1813.4-1246, along with H.E.S.S., HAWC, and LHAASO source extension. HESS J1813-126 significance map, as observed by H.E.S.S., is also shown in Fig. \ref{fig: morpho}. Note that no previously undetected source was found within H.E.S.S. extent and in the vicinity of the 4FGL source through the source finding algorithm in the phase-averaged standard analysis.

\begin{figure}
    \centering
    \includegraphics[width=\columnwidth]{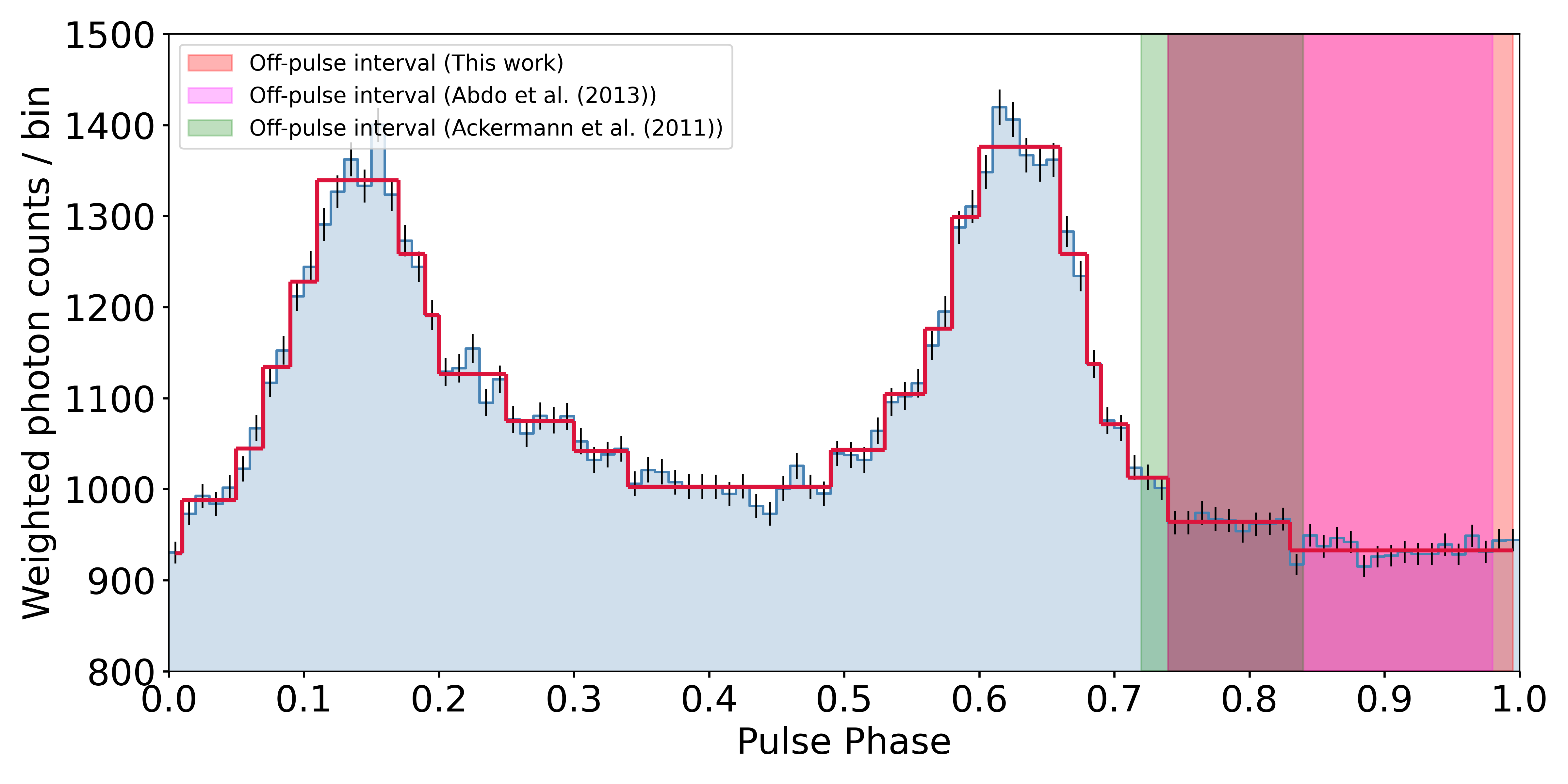}
    \caption{Weighted pulse profile of PSR J1813-1246 (4FGL J1813.4-1246) over one rotational cycle (pulse phase 0-1).\ It was constructed using 100 uniform phase bins per period from events between 0.1 and 500 GeV with the model weight from \cite{bruel19}. The blue histogram shows the weighted photon counts per phase bin, while the solid red line indicates the Bayesian block decomposition of statistically significant structures in the weighted pulse profile. The red shaded region signifies the off-pulse interval estimated from this work, whereas the magenta and green shaded regions signify the same obtained in \cite{abdo13} and \cite{ackermann11}, respectively.}
    \label{fig: phase_histogram}
\end{figure}

Next, we discuss the result from the pulsar phased analysis, the method of which has also been discussed in Appendix \ref{lat}.
Before the analysis, we constructed weighted pulse phase histograms and applied the Bayesian blocks algorithm \citep{jackson05,scargle13}, which adaptively identifies significant changes in photon count rate. This yielded an off-pulse interval (0.74–0.99) where pulsar contamination should be minimal. Note that our off-peak definition closely matches that given in \citet[0.74 - 0.98]{abdo13}; however, it is somewhat different in \citet[0.72 - 0.84]{ackermann11}. Further note that \cite{abdo13} has already cautioned that systematic uncertainties in discerning magnetospheric and diffuse background emission may complicate the detection of faint off-peak GeV emission, and additionally, some residual low-altitude magnetospheric emission may still persist. Using a larger dataset, we reexamined the source to search for any residual off-peak emission that could potentially be crucial evidence in favor of HESS J1813-126 being a GeV PWN. The weighted pulse phase histogram, along with the off-pulse phase selections from this work and previous literature \citep{ackermann11,abdo13}, are shown in Fig. \ref{fig: phase_histogram}, where the model weight has been considered from \cite{bruel19}. Note that the narrower phase interval of 0.83 - 0.99, identified through the Bayesian blocks analysis, was not adopted as the off-pulse phase region, since the limited photon statistics in this interval were insufficient to yield a statistically significant off-pulse emission.

During the off-pulse phase of the GeV counterpart of the pulsar, 4FGL J1813.4-1246, a new source, PS J1813.3-1246, was detected at a minor offset of 0.011$^ {\circ}$ from the position of the pulsar. The source was detected at the {best-fit localization of $\mathrm{RA} = 273.3459^{\circ} \pm 0.0122^{\circ}$ and $\mathrm{Dec} = -12.7781^{\circ} \pm 0.0120^{\circ}$ with a detection significance of $23.8\sigma$.} The PS source is consistent with the containment radii of the sources detected by H.E.S.S., HAWC, and LHAASO. The spectrum of the source is explained by a simple power law, which has {best-fit parameters of $\alpha_{PL} = -3.123 \pm 0.001$ at a reference energy scale of $E_{\mathrm{0, PL}} = 1000 \ \mathrm{MeV}$.}
{The energy flux integrated within 1-500 GeV was found to be $(2.419 \pm 0.122)\times10^{-11}~\mathrm{erg~cm^{-2}~s^{-1}}$.} We further examined the spatial extension of the source and {found that the \texttt{RadialDisk} model yields a best-fit extension of $0.0872^{+0.0236}_{-0.0316}\,\mathrm{deg}$ with $\mathrm{TS_{ext}} \approx 2.8$ ($\sim 1.7\sigma$), 
while the \texttt{RadialGaussian} model gives $0.0752^{+0.0268}_{-0.0292}\,\mathrm{deg}$ with $\mathrm{TS_{ext}} \approx 2.4$ ($\sim 1.5\sigma$), 
thus clearly hinting at a point-like morphology.} 

Given that the off-pulse source spectrum is adequately described by a simple power law and no significant
extension is detected, the balance of evidence favors residual pulsar magnetospheric contamination rather than a GeV PWN. First, the source is point-like and positionally coincident with the pulsar, whereas a nebular IC component is expected to be spatially extended at GeV energies. Second, the measured power law index is relatively soft for a GeV PWN ($\sim 1.8-2.2$) and does not connect naturally to the TeV spectrum without introducing a break, but it is compatible with an unpulsed or tail magnetospheric component leaking into the off-pulse window due to imperfect gating or ephemeris uncertainties. Third, previous studies report no significant off-pulse GeV emission \citep{ackermann11, abdo13, marelli14}, consistent with our finding that any putative
nebular signal would be subdominant. We therefore interpret the off-pulse SED as an
upper limit and attribute the detected GeV flux to residual
magnetospheric radiation rather than a GeV nebular emission. A future \textit{Fermi}-LAT phased analysis with an updated ephemeris will be crucial to search for any GeV emission hidden beneath the pulsar signal. Figure~\ref{fig: morpho} also shows the significance map of the off-pulse source PS~J1813.3-1246.

\begin{figure*}
    \centering
    \includegraphics[width=0.24\textwidth]{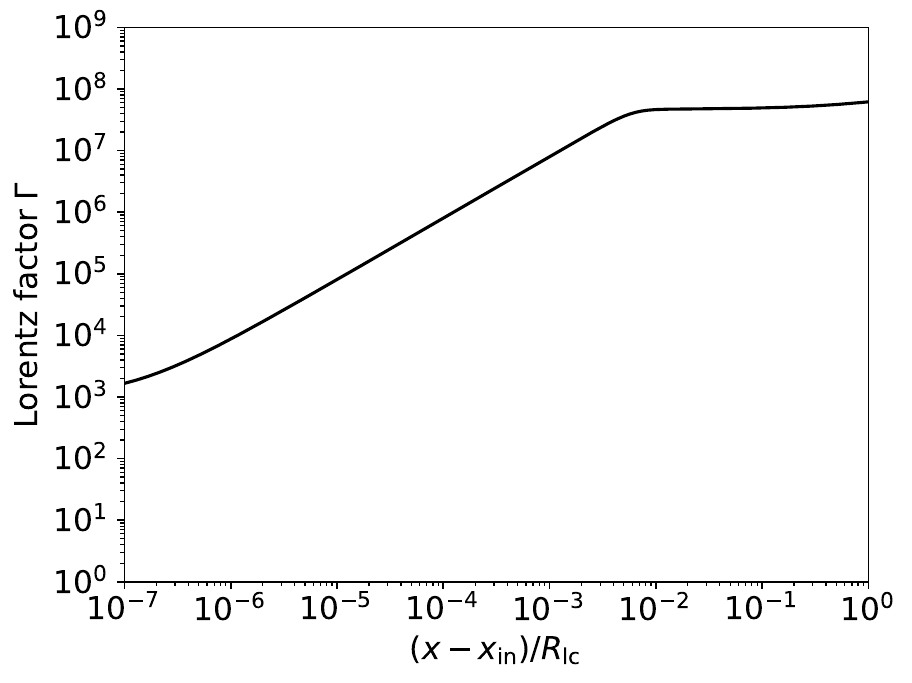}\hfill
    \includegraphics[width=0.24\textwidth]{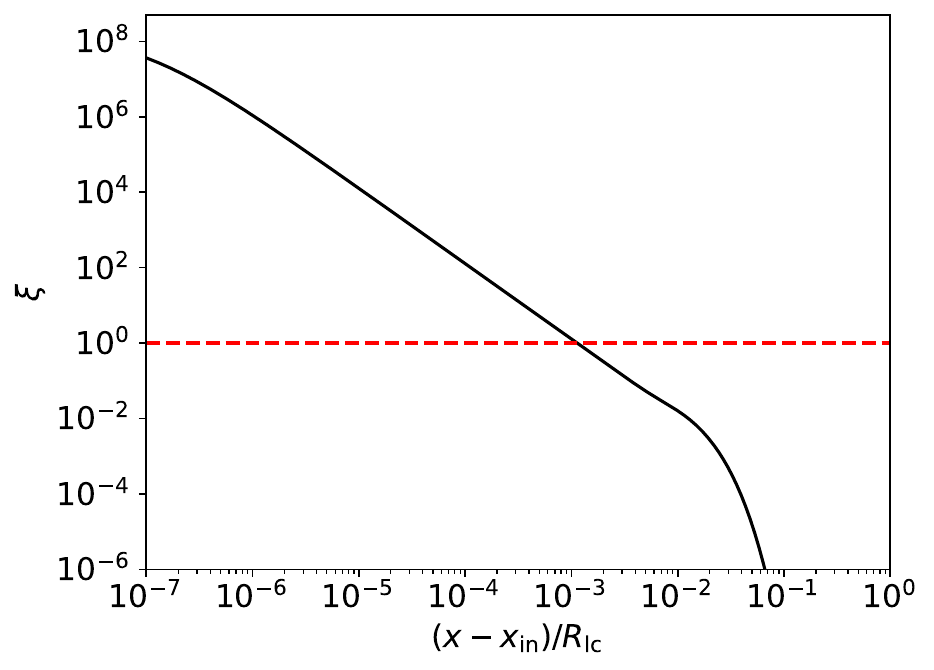}\hfill
    \includegraphics[width=0.24\textwidth]{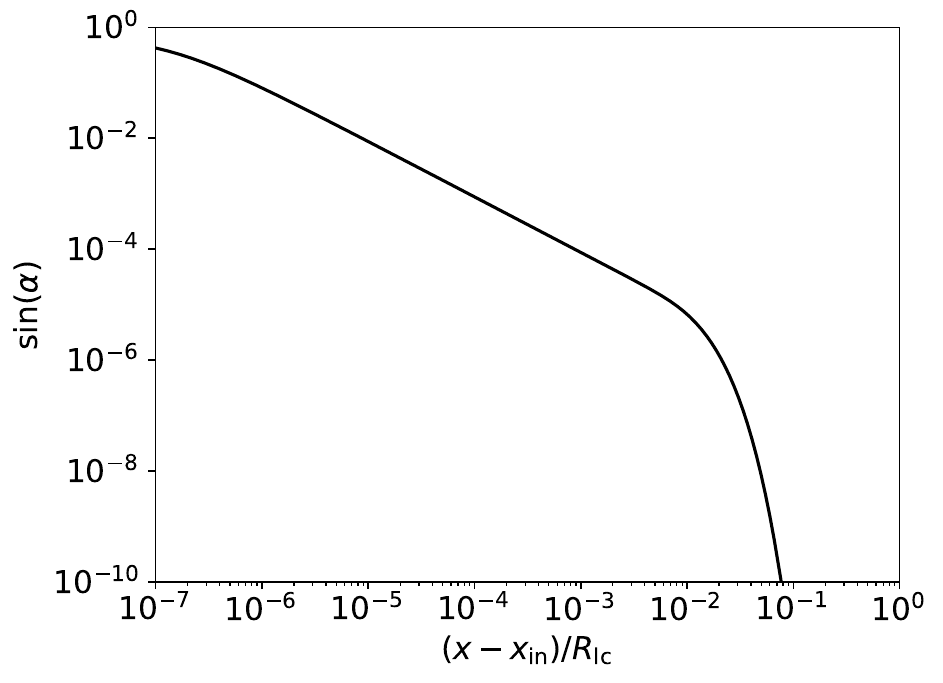}\hfill
    \raisebox{0.025\height}{\includegraphics[width=0.25\textwidth]{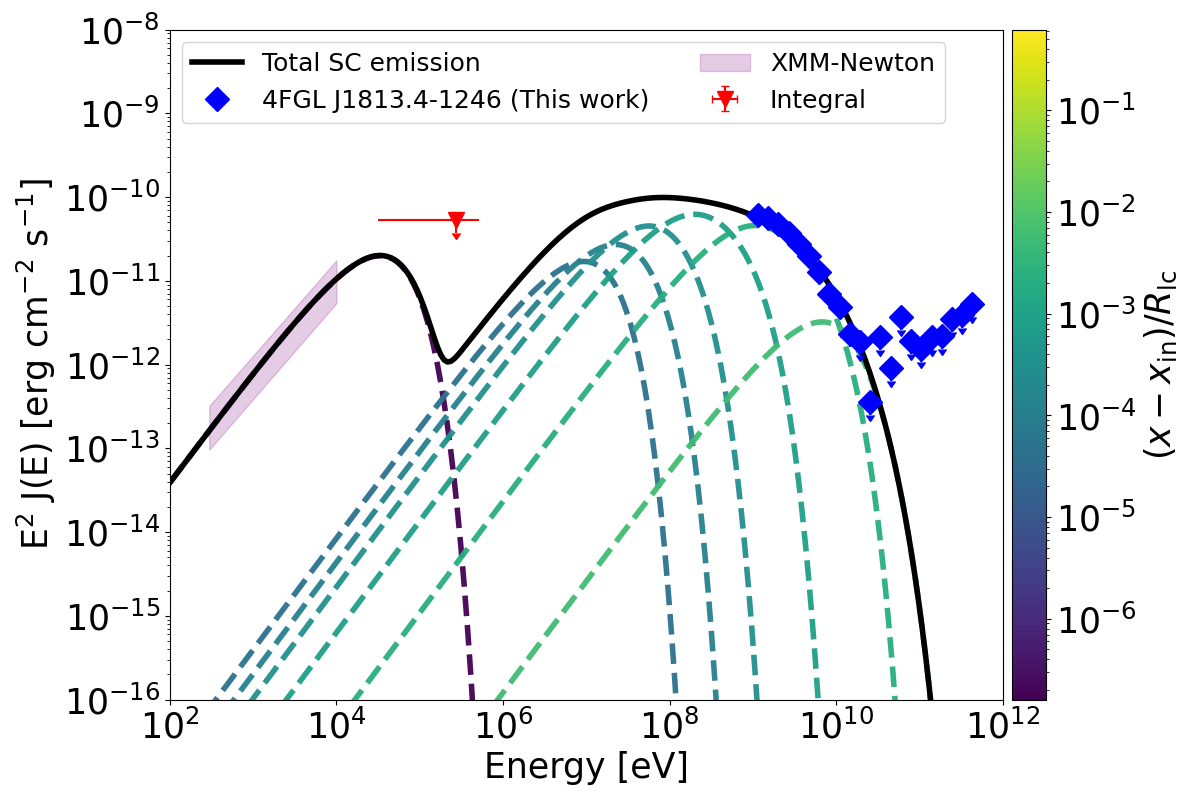}}
    \caption{Distance variation of different SC model parameters and the spectrum of PSR J1813-1246: Lorentz factor ($\Gamma$; first panel), SC parameter ($\xi$; second panel), and pitch angle ($\mathrm{sin} \ \alpha$; third panel). In the second panel, the dashed red line marks $\xi = 1$, i.e., the transition point between synchrotron- and curvature-dominated regimes. In the last panel, the model spectrum is plotted against the observed SED obtained from the \textit{Fermi}-LAT data analysis of 4FGL J1813.4-1246 and the X-ray SED and upper limit from \textit{XMM-Newton} and INTEGRAL data, respectively, as obtained from \cite{marelli14} and \cite{guevel25}. In the last panel, decomposed emission from different parts of the emitting region is also shown; the colorbar signifies the distance from the inner boundary ($x_{\rm in}$) in units of the light-cylinder radius ($R_{\rm lc}$).}
\label{fig: psr_char}
\end{figure*}

Next, we move on to describe the gamma-ray and X-ray SEDs of PSR~J1813-1246 with the SC framework. In this picture, relativistic $e^\pm$ pairs are accelerated by a parallel electric field $E_\parallel$ in the outer pulsar magnetosphere and radiate while both spiraling around and sliding along curved magnetic field lines. The SC formalism naturally interpolates between the synchrotron and curvature limits and has been shown to reproduce the curved spectra observed by {\it Fermi}–LAT in many energetic pulsars. We briefly present the SC model equations in Appendix \ref{pulsar} \citep[for a more detailed discussion, see, e.g.,][]{zheng96,vigano15b,vigano15c,vigano15,vigano15d,torres18,torres19, desarkar24}. 
The gamma-ray SED corresponds to the \textit{Fermi}-LAT source 4FGL~J1813.4-1246 (this work), whereas the \textit{XMM-Newton} X-ray spectrum and INTEGRAL X-ray upper limit are taken from \citet{marelli14} and \cite{guevel25}. 
Assuming a minimum pulsar distance of $r_s = 2.5$~kpc, {the SC parameters that reproduce the data are 
$\mathrm{log}(N_0) = 32.3$, 
$\log(E_\parallel/\mathrm{V\,m^{-1}})=9.25$, 
magnetic field gradient $b_{sc} = 3.0$, 
and the scale length $\mathrm{log}(x_0/R_{\mathrm{lc}}) = -3.4$.} 
The resulting model SED, together with the observational data, is shown in Fig. \ref{fig: psr_char}, along with the Lorentz factor $\Gamma$, the SC parameter $\xi$, and the pitch angle $\sin\alpha$ as functions of the distance from the inner boundary $x_{\mathrm{in}}$ in units of the light-cylinder radius $R_{\mathrm{lc}}$. 
In the SED plot of Fig. \ref{fig: psr_char}, the decomposed emission from different parts of the emitting region has also been shown, as a function of distance from $x_{\mathrm{in}}$ in units of $R_{\mathrm{lc}}$.
The model reproduces both bands well: the synchrotron component explains the X-ray SED, while the curvature component accounts for the GeV spectrum.
The parameter $\xi$ acts as a diagnostic of the dominant regime: if $\xi \ll 1$, curvature radiation prevails, whereas $\xi \gg 1$ signals synchrotron dominance. 
From Fig.~\ref{fig: psr_char} we see that the transition between the two regimes occurs at around $(x - x_{\mathrm{in}})/R_{\mathrm{lc}} \approx 10^{-3}$, suggesting that the X-rays are produced at low altitudes where synchrotron losses dominate, while the gamma rays are emitted further out along the field lines, where curvature radiation becomes the main channel.

It is to be noted that the origin of the observed quarter phase lag between the X-ray and gamma-ray light curves of PSR J1813-1246 \citep{marelli14} remains a puzzling issue. Several possibilities could, in principle, account for such a lag, including emission from spatially distinct regions within the gap, such as a synchrotron-dominated inner zone and a curvature-dominated outer zone, and a presence of strong magnetic field sweepback near the light-cylinder \citep{contopoulos99, spitkovsky06, kalapotharakos12, dyks03, dyks08, dyks04, dyks04a, watters09, bai10}, which can shift the formation of caustic peaks. However, quantitatively addressing these possibilities requires a fully self-consistent description of the magnetospheric electrodynamics, ideally within global resistive magnetohydrodynamic or kinetic (particle-in-cell) frameworks similar to that discussed in \cite{marelli14}, or a more methodological approach \citep[see, e.g.,][]{vigano19, pascual24} that incorporates realistic field topology, particle acceleration, and radiative transport in a coupled manner. Such an analysis falls beyond the scope of the present work, which is primarily focused on elucidating the pulsar halo origin of HESS J1813–126 and its connection to PSR J1813–1246. A detailed investigation of the phase-lag problem under the SC framework is therefore left to future dedicated studies.

\begin{figure*}
    \centering
    \includegraphics[width=0.31\textwidth]{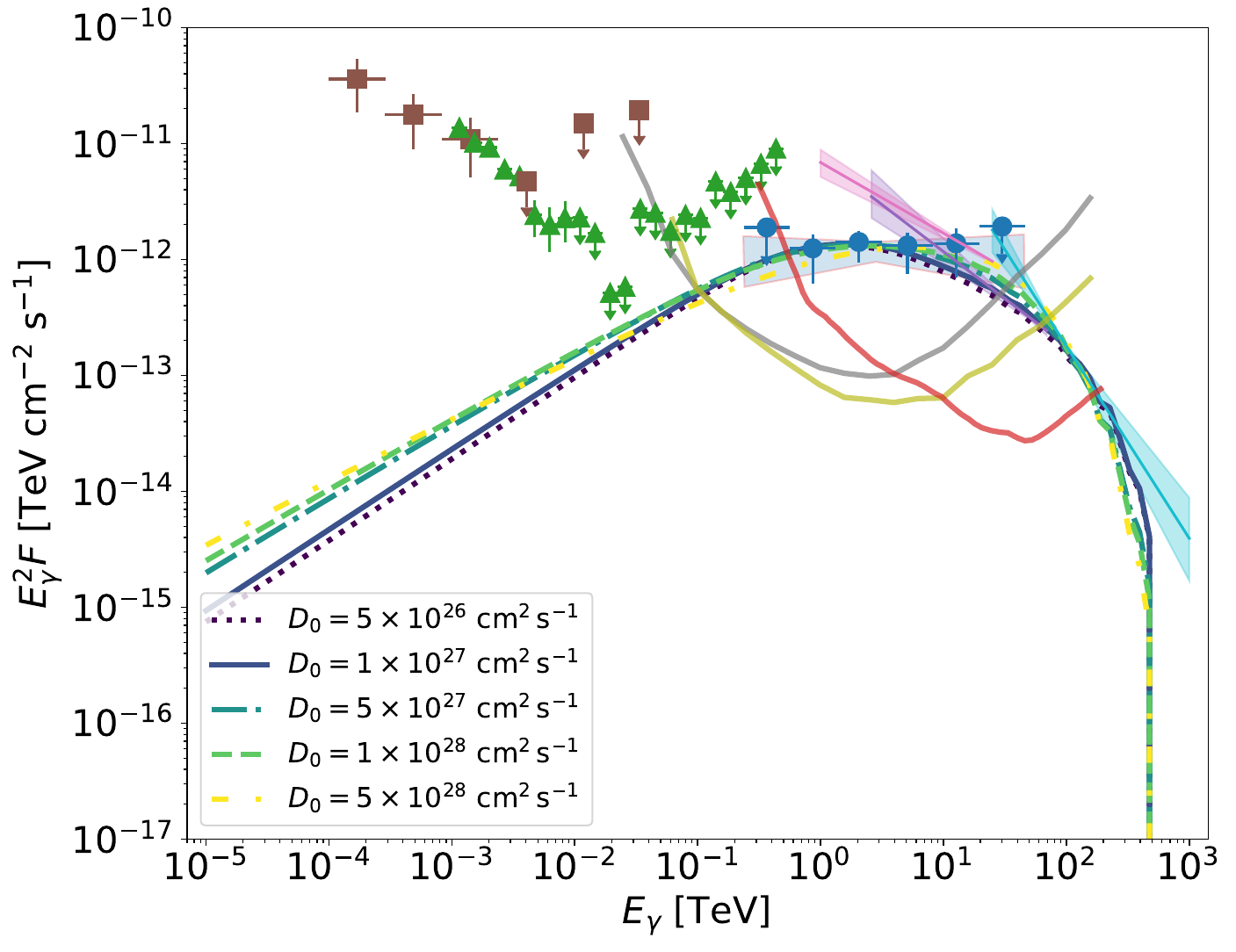}
    \includegraphics[width=0.31\textwidth]{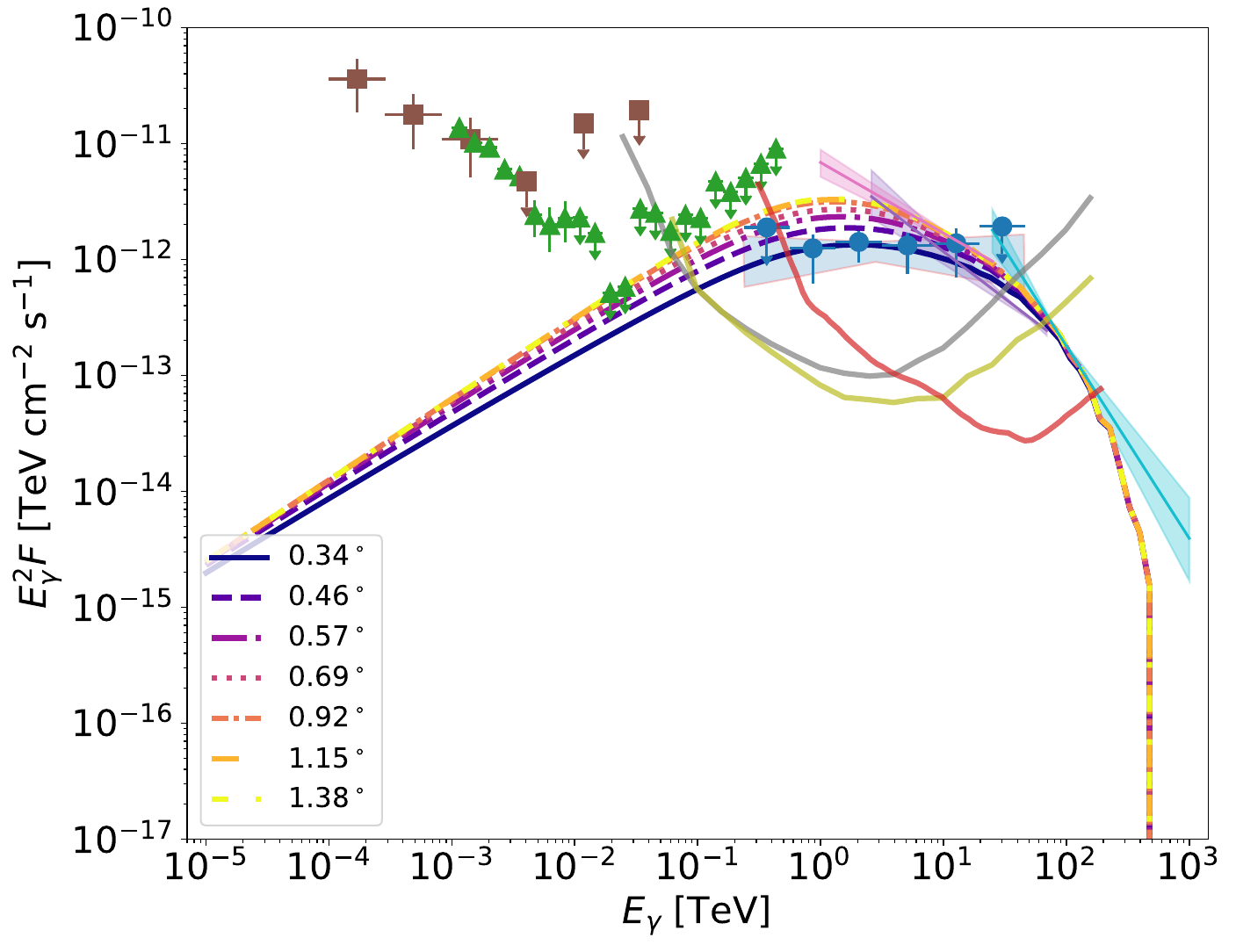}
    \includegraphics[width=0.3\textwidth]{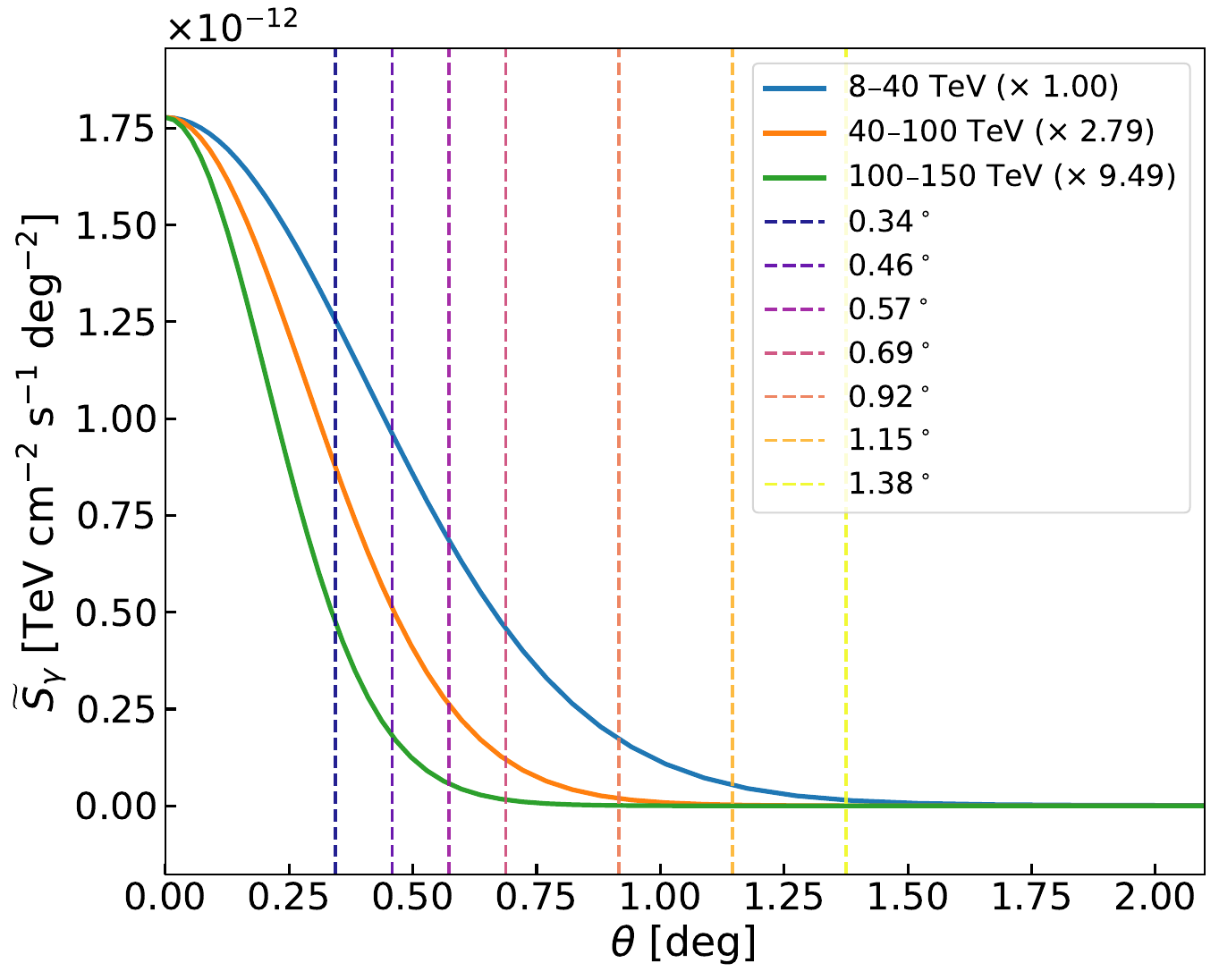}
    \caption{Exploration of SEDs and SBPs of HESS J1813-126 in the 2ZISD framework. Left: SEDs corresponding to different choices of inner zone diffusion coefficients. The $D_0$ in the plot signifies the normalization of the inner zone diffusion coefficient. Middle: Angle-integrated SEDs with increasing aperture sizes starting from 0.34$^{\circ}$, assuming $D_{\mathrm{0}} = 5 \times10^{27} \ \mathrm{cm^2 \ s^{-1}}$. In these two panels, the brown squares and green triangles (along with the upper limits) correspond to the off-pulse emission from \cite{ackermann11} and this work (PS J1813.3-1246), respectively. The blue data points and butterfly plot correspond to the H.E.S.S. data, whereas the purple, pink, and sky blue butterfly plots correspond to the HAWC, LHAASO WCDA, and KM2A observations, respectively. We also provide the sensitivity curves for the CTA-North (gray), CTA-South (yellow-green), and SWGO (red) for comparison. Right: SBPs for three different energy ranges scaled to start from the same point. Vertical lines correspond to the aperture sizes and have the same color scheme as in the middle panel.  }
    \label{fig: halo_SED_2ZISD}
\end{figure*}

\begin{table*}[t]
\centering
\caption{Parameters adopted for the 2ZISD model.}
\begin{tabular*}{\textwidth}{@{\extracolsep{\fill}}ccccc}
\hline\hline
$D_0$ (at 100 TeV) [cm$^2$\,s$^{-1}$] & $r_{\rm 2z}$ (pc) & $\Xi_{\rm 2z}$ & $\eta$ & $E_{\mathrm{e,c}}$ (TeV) \\
\hline
$5 \times 10^{26}$ & 50 & 0.0015   & 0.004 & 400 \\
$1 \times 10^{27}$ & 50 & 0.0031   & 0.005  & 350 \\
$5 \times 10^{27}$ & 50 & 0.0158   & 0.010  & 170 \\
$1 \times 10^{28}$ & 50 & 0.0316   & 0.016  & 140 \\
$5 \times 10^{28}$ & 50 & 0.1582    & 0.060  & 100 \\

\hline
\end{tabular*}

\tablefoot{  $D_0$ signifies the normalization of the inner zone diffusion coefficient.}\label{tab:2zone_params}
\end{table*}

Finally, we explored the pulsar halo origin of HESS J1813-126. To describe the transport of $e^\pm$ pairs and to further discuss the SED and morphology of the resulting gamma-ray emission via the IC upscattering of the cosmic microwave background (CMB), infrared photons from interstellar dust, and optical stellar fields, in this section we consider three scenarios that have been proposed in the literature: the two-zone isotropic suppressed diffusion (2ZISD) model, the ballistic-to-diffusion (B2D) model, and the anisotropic diffusion (AD) model. These models differ mainly in how the spatial diffusion of particles is parameterized, but share the same general transport framework. We briefly discuss the formulae pertinent to the transport framework in Appendix \ref{halo} \citep[for a more detailed discussion, see][]{fang25}. We applied all of these three models to explore the pulsar halo origin of HESS J1813-126 using the \texttt{PHECT} code \citep{fang25}. 

\begin{figure*}
    \centering
    \includegraphics[width=\columnwidth]{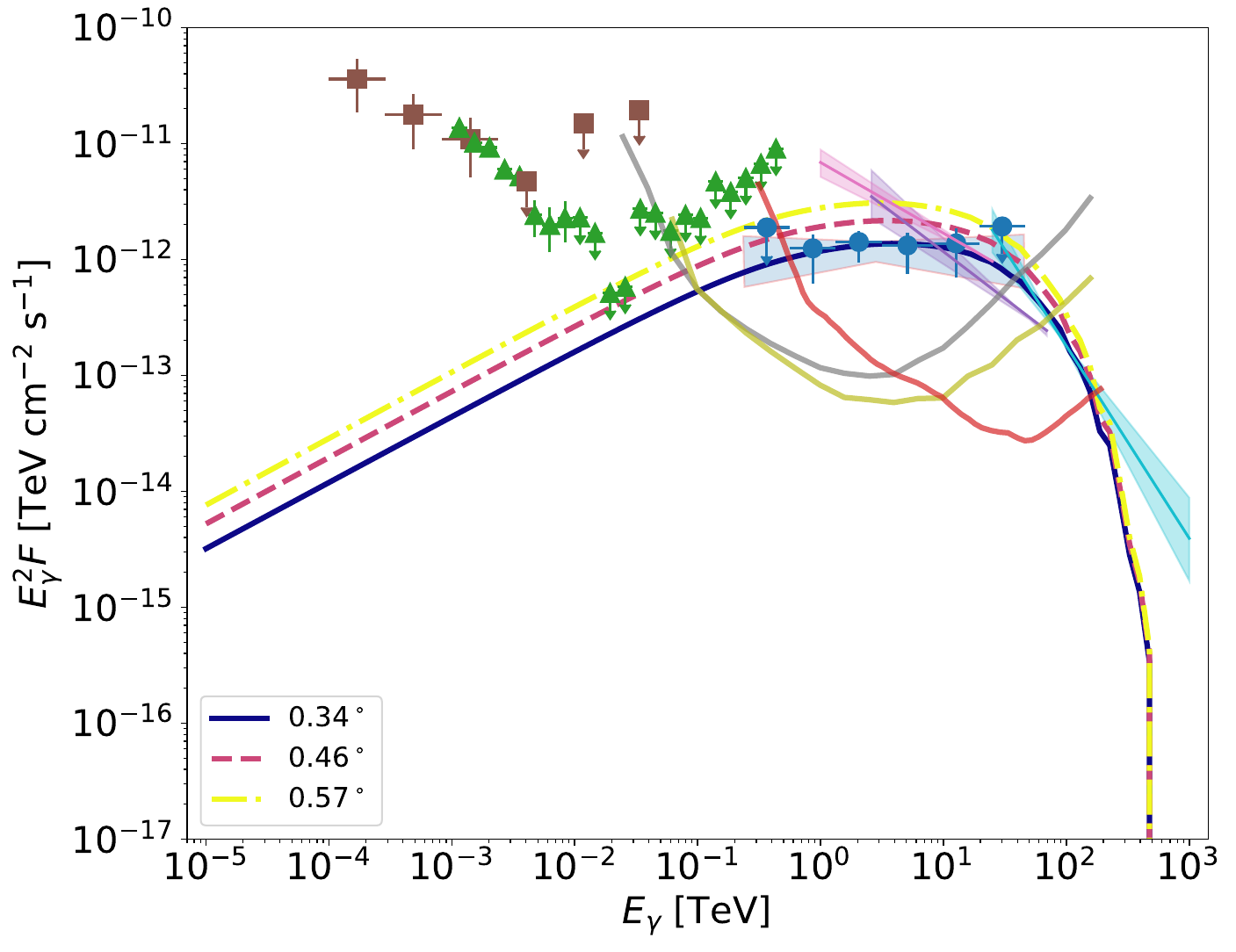}
    \includegraphics[width=0.95\columnwidth]{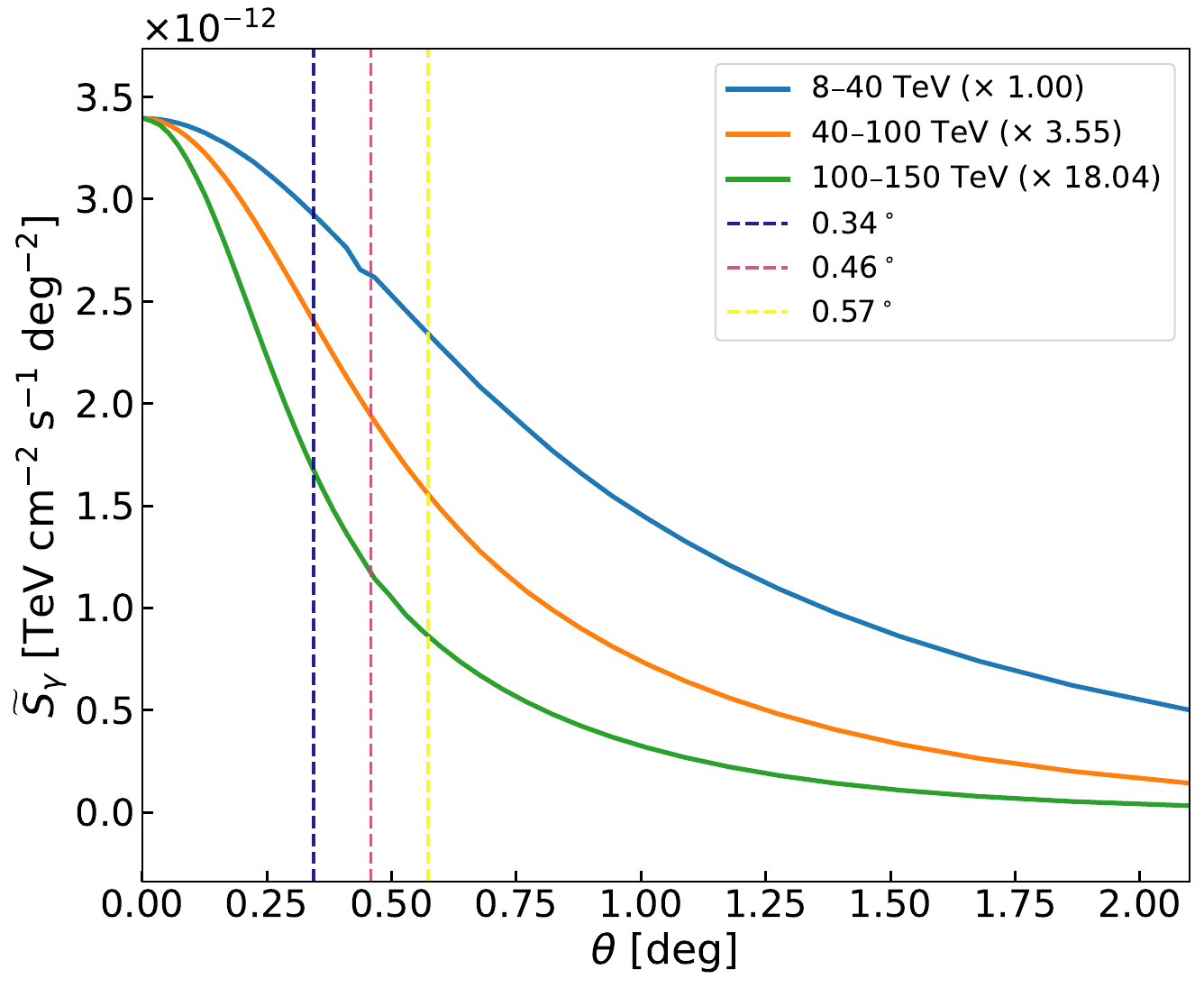}
    \caption{Exploration of the SEDs and SBPs of HESS J1813-126 in the B2D framework. Left: Angle-integrated SEDs, calculated for increasing aperture sizes, together with the same data points, upper limits, butterfly plots, and sensitivity curves as in Fig.~\ref{fig: halo_SED_2ZISD}, using the same color scheme. Right: SBPs for three different energy ranges scaled to start from the same point. Vertical lines correspond to the aperture sizes and follow the same color scheme as in the left panel.}
    \label{fig: halo_SED_B2D}
\end{figure*}

\begin{figure*}
    \centering
    \includegraphics[width=0.33\textwidth]{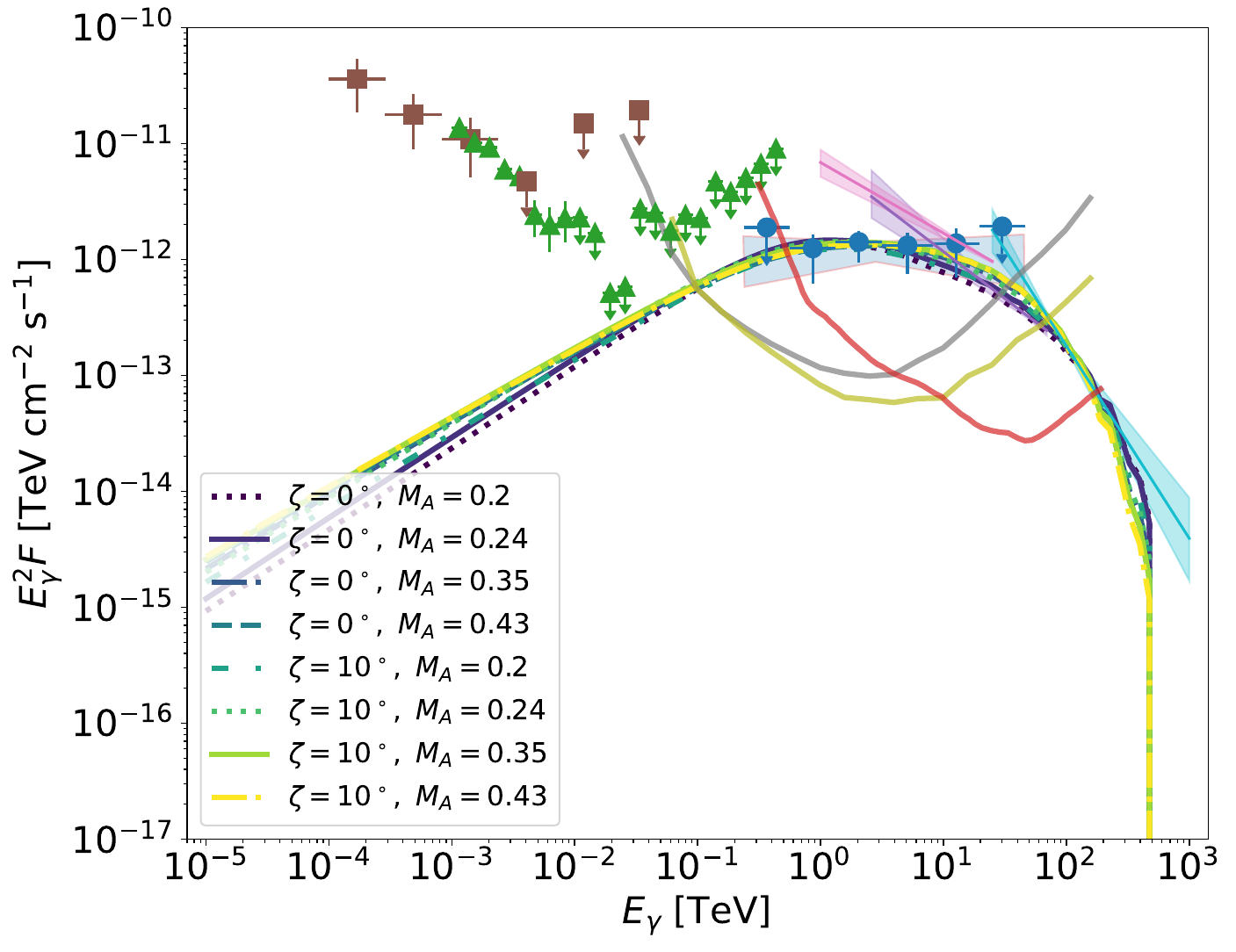}
    \includegraphics[width=0.33\textwidth]{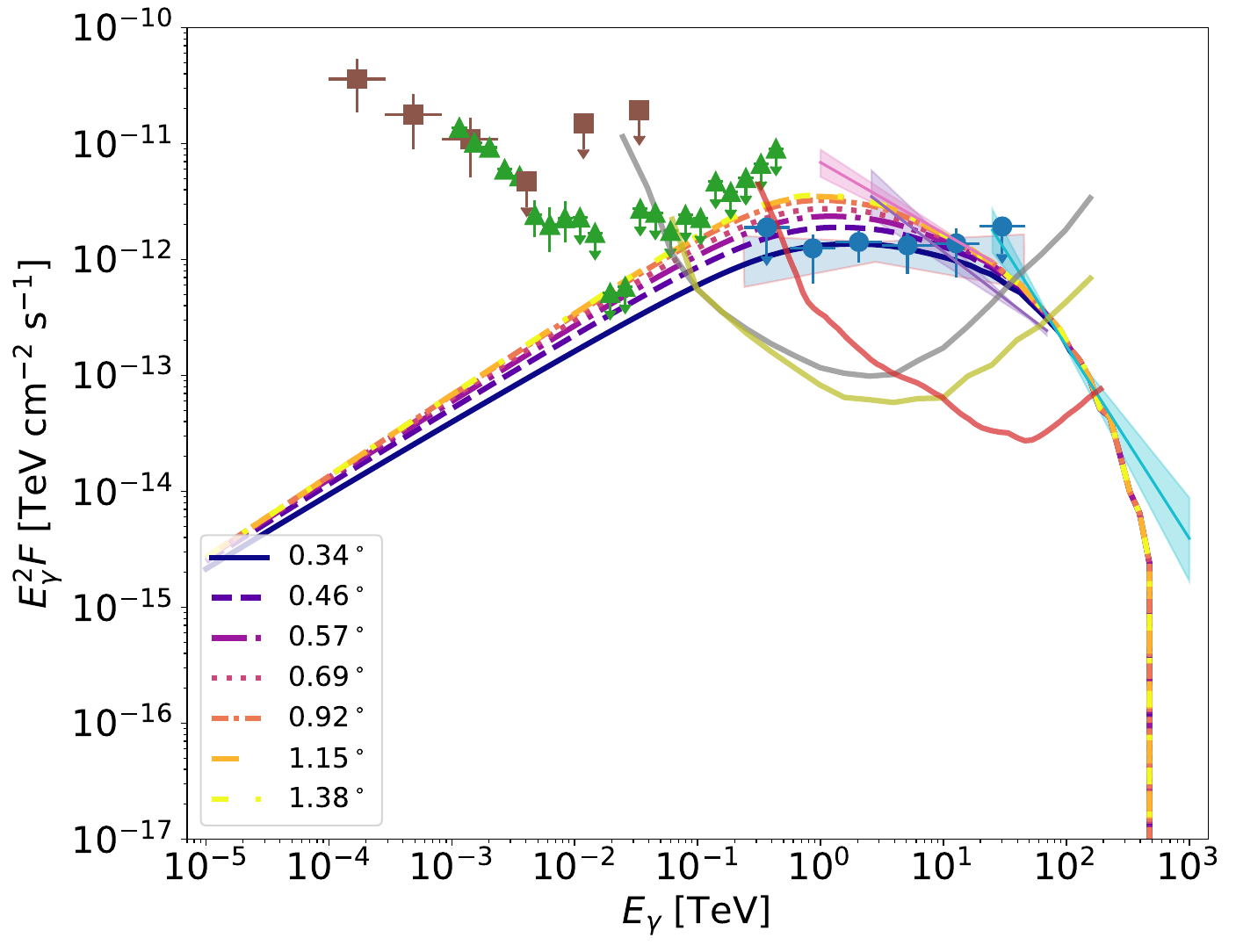}
    \includegraphics[width=0.33\textwidth]{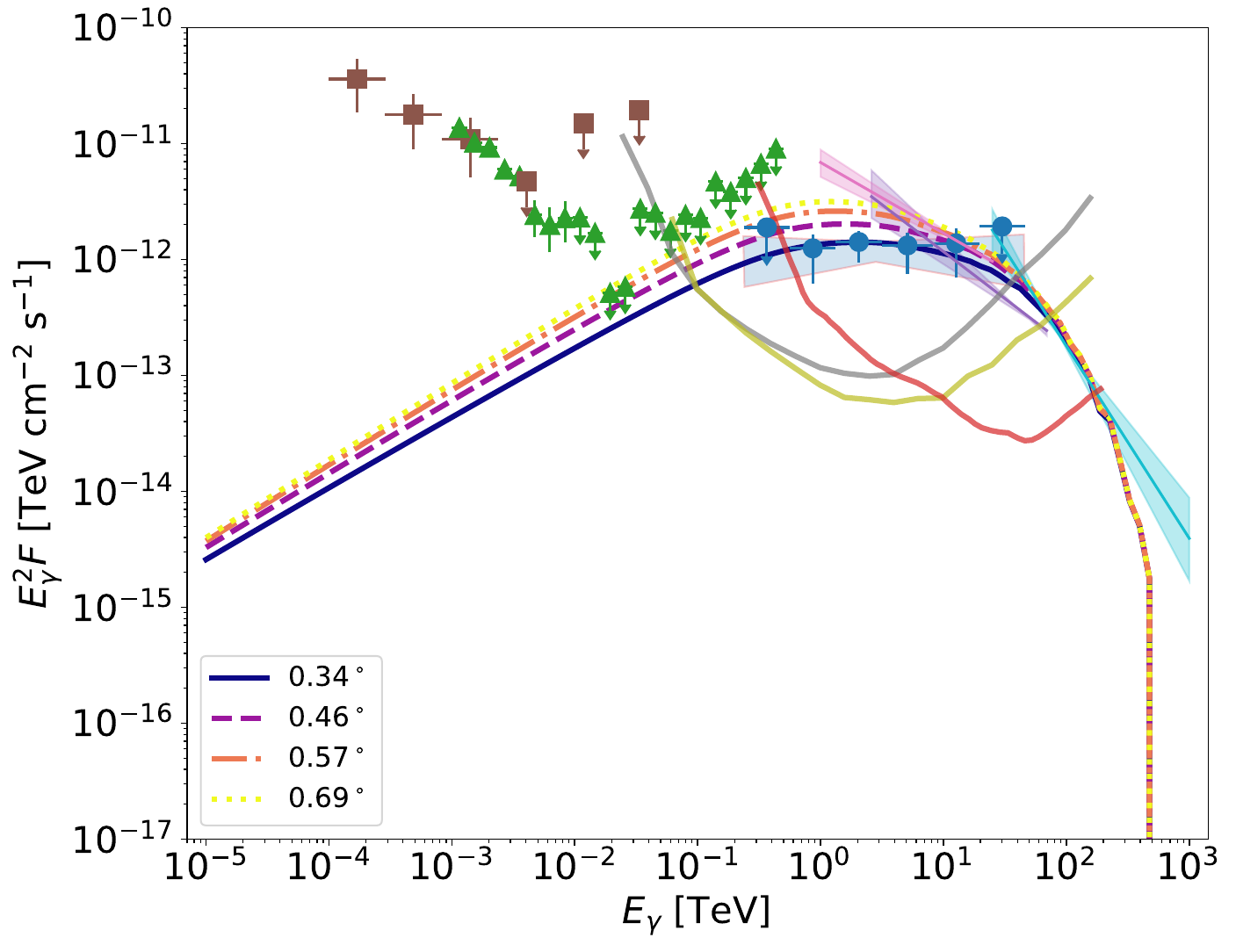}
    \caption{Exploration of SEDs of HESS J1813-126 in the AD framework. Left: Model SEDs of HESS J1813-126 
    for different $M_A-\zeta$ combinations. Middle and right: SEDs for the combinations $M_A = 0.35$ and $\zeta = 0^{\circ}$, and $M_A = 0.35$ and $\zeta = 10^{\circ}$, respectively. In both panels, the primary SEDs correspond to an aperture of 
    $0.34^{\circ}$, while additional SEDs show progressively larger apertures. The same data points, upper limits, butterfly plots, and sensitivity curves are provided as in Fig.~\ref{fig: halo_SED_2ZISD}, using the same color scheme.}
    \label{fig: halo_SED_AD_1}
\end{figure*}

\begin{figure*}
    \centering
    \includegraphics[width=0.9\textwidth]{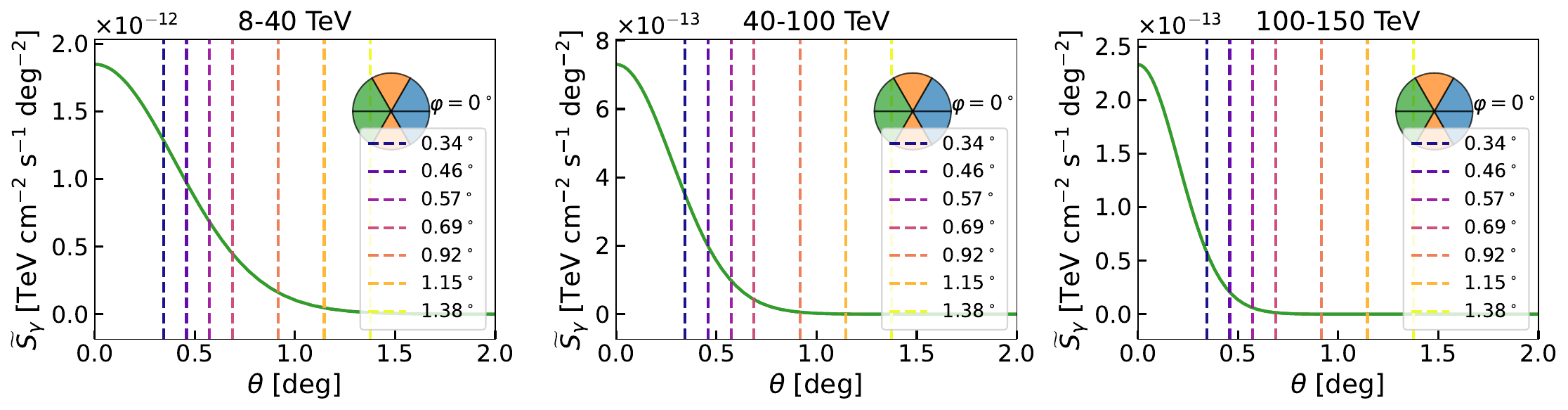}
    \includegraphics[width=0.9\textwidth]{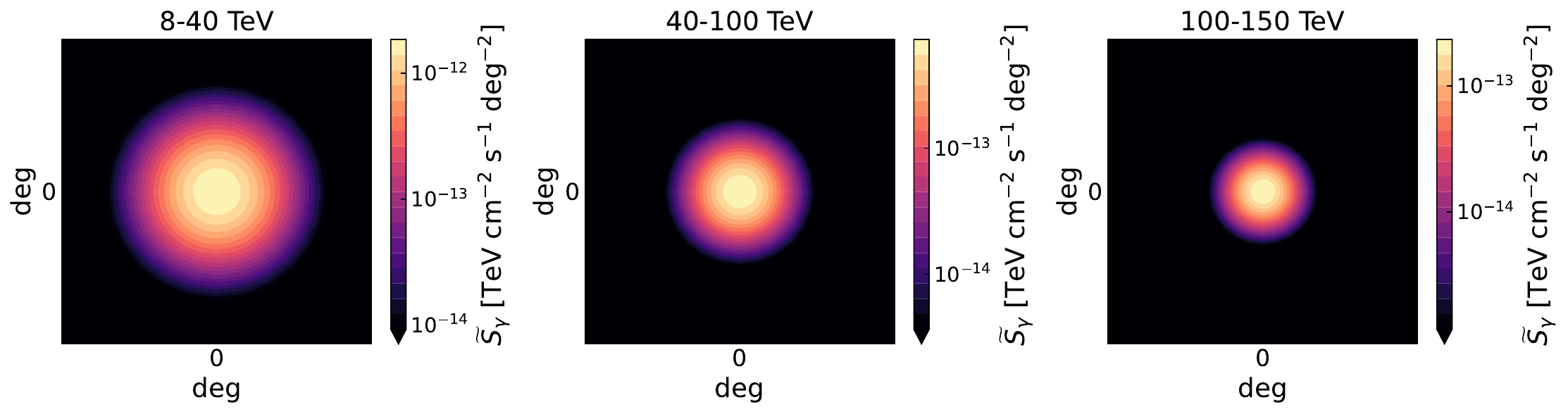}
    \includegraphics[width=0.9\textwidth]{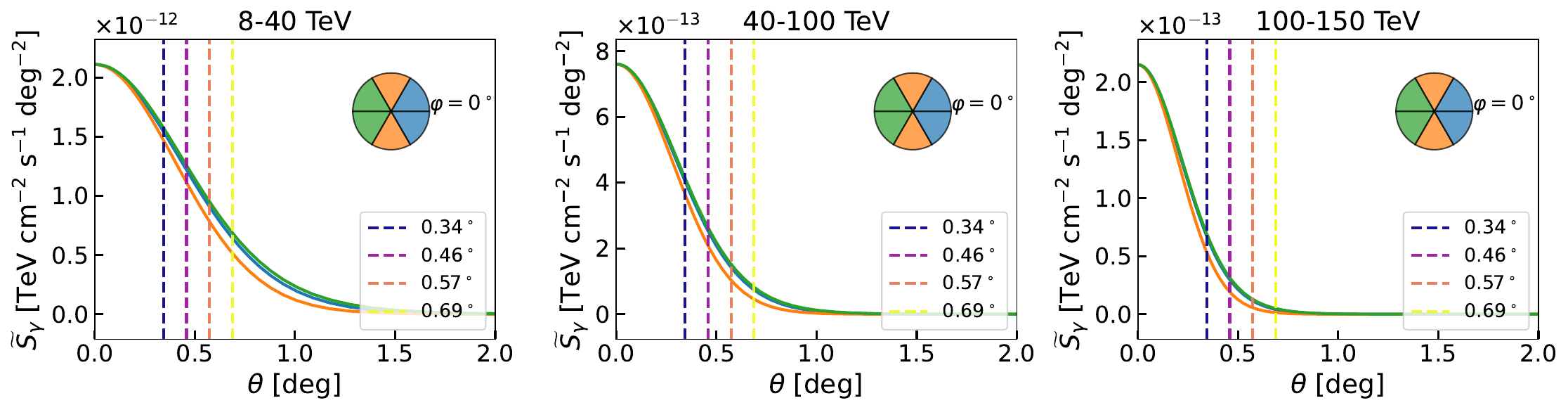}
    \includegraphics[width=0.9\textwidth]{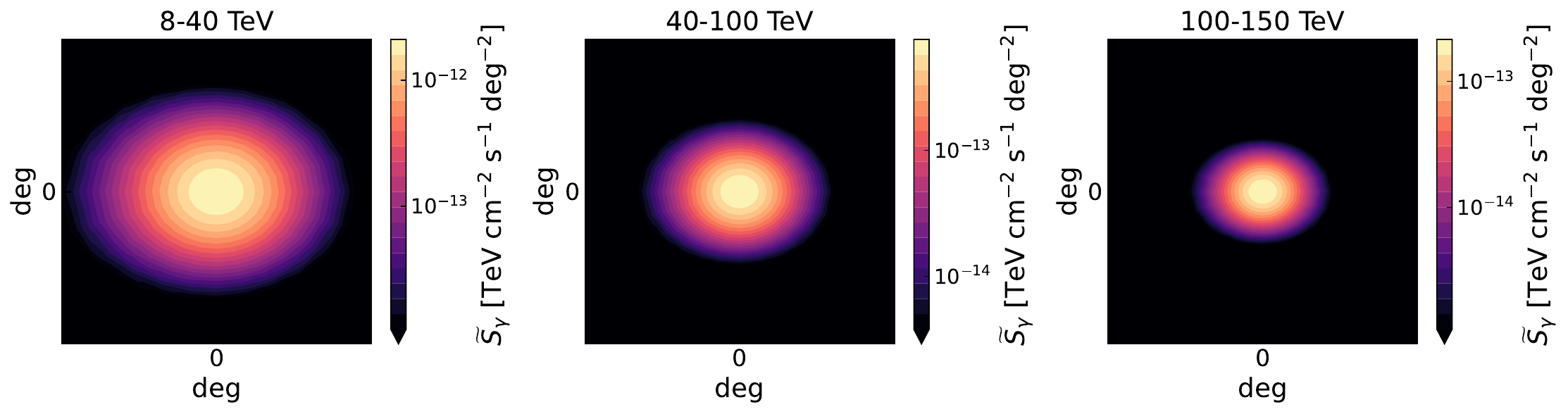}
    \caption{Exploration of SBPs and gamma-ray maps of HESS J1813-126 in the AD framework. The model SBPs are shown for two combinations of $M_A$ and $\zeta$ and for three energy ranges, scaled to start from the same point, and along with vertical lines, corresponding to the same aperture sizes and with the same color scheme as in Fig. \ref{fig: halo_SED_AD_1}. The SBP curves are divided across different azimuthal intervals, visualized as circular pie charts in the corresponding plots. The figure also shows gamma-ray maps in three energy ranges for the two considered combinations, to emphasize the anisotropy of the AD model. Top two panels: Results for $M_A = 0.35$ and $\zeta = 0^{\circ}$. Bottom two panels: Results for $M_A = 0.35$ and $\zeta = 10^{\circ}$. }
    \label{fig: halo_SED_AD_2}
\end{figure*}

In this paper we primarily focus on explaining the gamma-ray SED using IC emission, and do not delve into explaining the X-ray upper limit with synchrotron emission. \cite{guevel25} has reported no evidence of excess X-ray emission, and from the calculated X-ray upper limits (4.32 $\times 10^{-4}$ keV$^{-1}$ cm$^{-2}$ s$^{-1}$ and 5.38 $\times$ 10$^{-4}$ keV$^{-1}$ cm$^{-2}$ s$^{-1}$ at 1 keV assuming $E^{-2}$ spectrum), they have posited that the magnetic field in the region should be in the range 3.5-6.5 $\mu$G, implying that the halo magnetic field is not significantly enhanced compared to the average Galactic magnetic field. In this work, we used a representative halo magnetic field of 5 $\mu$G to be consistent with \cite{guevel25}. The current age of the system is unknown, so we considered it to be $t_{\mathrm{s}} = 30 \ \mathrm{kyr}$. We chose this value to be consistent with \cite{albert25}, who posited that the typical age of pulsars that contain a pulsar halo is $> 20 \ \mathrm{kyr}$. The initial spin-down timescale can be calculated as $\tau_0 = \tau_c - t_{\mathrm{s}} \approx 13 \ \mathrm{kyr}$, given that we considered $n = 3$. The spectral index of the particle spectrum is chosen to be $p = 1.6$ and $s = 1$, making it a typical power law with exponential cutoff spectrum. Additionally, to calculate the IC emission, we considered radiation fields from CMB, interstellar dust, and starlight, with the parameters: $T_{\mathrm{CMB}} = 2.73 \ \mathrm{K}$, $\epsilon_{\mathrm{CMB}} = 0.26 \mathrm{\ eV \ cm^{-3}}$, $T_{\mathrm{dust}} = 20 \ \mathrm{K}$, $\epsilon_{\mathrm{dust}} = 0.3 \mathrm{\ eV \ cm^{-3}}$, $T_{\mathrm{starlight}} = 5000 \ \mathrm{K}$, and $\epsilon_{\mathrm{starlight}} = 0.3 \mathrm{\ eV \ cm^{-3}}$, respectively. The distance of the source was kept the same as the pulsar at 2.5 kpc. Finally, to calculate the intrinsic surface brightness profiles (SBPs) of different models, we convolved the model predictions with a symmetric 2D Gaussian point spread function (PSF; see Eq. \ref{eq:PSF} and the discussion in Appendix \ref{halo}). The adopted width $\sigma_{\rm PSF}$ in the figures is illustrative, serving only to demonstrate the effect of instrumental smearing, and should not be taken as corresponding to any specific instrument response. In practice, $\sigma_{\rm PSF}$ can be set to the appropriate value for a given telescope. For this work, as a representative case, we considered the width of the 2D Gaussian PSF as 0.35$^{\circ}$ (in the 8-40 TeV range), 0.25$^{\circ}$ (in the 40-100 TeV range), and 0.18$^{\circ}$ (in the 100-150 TeV range). The primary free parameters varied to explain the gamma-ray data for the three models are the conversion efficiency ($\eta$) and the cutoff energy ($E_{e, c}$). Furthermore, the inner zone diffusion coefficient in the 2ZISD model, and different combinations of $M_A$ and $\zeta$ in the AD model have been explored. In the B2D model, the diffusion coefficient is fixed at the standard ISM value, i.e., $D_{\rm ISM, 0} = 3.16 \times 10^{29} \ \mathrm{cm^2 \ s^{-1}}$ at 100 TeV, and Kolmogorov-like $\delta = 0.3$.
Figures~\ref {fig: halo_SED_2ZISD}--\ref{fig: halo_SED_AD_2} show the modeled SEDs and  SBPs of HESS~J1813-126 under the three transport models introduced in Appendix \ref{halo}. In the SED panels of all figures, the primary curve corresponds to an aperture of $0.34^\circ$ (the upper limit of integration in Eq. \ref{eq: theta_integrated_spectrum}), matching the spectral extraction region measured by H.E.S.S. The additional curves show progressively larger apertures, which allow us to track how the integrated flux depends on the chosen extraction region. The considered apertures have also been indicated in the SBPs using vertical lines.

In Fig. \ref{fig: halo_SED_2ZISD} we show the results of the 2ZISD model. Given that the diffusion coefficient of the inner zone is unknown and cannot yet be constrained due to a lack of observed SBP data, we considered five different values of it that aptly cover the range of values typically considered in pulsar halo models. The SEDs, corresponding to these five choices, and integrated within $0.34^\circ$, reproduce the H.E.S.S., HAWC, and LHAASO data well. The SEDs also do not exceed the upper limit provided by $\textit{Fermi}$-LAT analysis. The required values of free parameters are given in the Table \ref{tab:2zone_params}. Then, we considered a representative inner diffusion coefficient value of $5 \times 10^{27}$ cm$^2$\,s$^{-1}$, consistent with that typically considered (see, e.g., \citealt{xi19}), and enlarged the aperture in multiple steps, finding that the SED primarily increases at the lower energies. This is understandable since the lower-energy electrons, having longer cooling timescales, can propagate farther from the pulsar before losing their energy, making the IC emission more spatially extended at lower energies. As a result, when a larger aperture is considered, more of this extended, lower-energy emission from the outer regions of the halo is collected, so the integrated SED increases significantly. From Fig. \ref{fig: halo_SED_2ZISD}, it can be seen that SEDs corresponding to enlarged apertures are consistent with the gamma-ray spectra posited by HAWC and LHAASO, before getting saturated. Enlarging the aperture even further does not increase the SED any more, since the steep decline of the particle density at the bubble boundary prevents much additional contribution.
The corresponding SBPs, shown in the right panel of Fig.~\ref{fig: halo_SED_2ZISD}, reinforce this picture. At 8--40~TeV, the profile exhibits a bright central peak followed by a radial decline, indicating that the bulk of the emission is confined within the inner region. As the photon energy increases, the SBPs become progressively narrower, reflecting the shorter cooling timescales of higher-energy electrons that prevent them from diffusing to large distances. The contrast between the extended low-energy profile and the compact high-energy profiles demonstrates the energy-dependent morphology that naturally arises in the 2ZISD framework. In practice, this implies that any aperture growth primarily enhances the flux in the lower-energy bands, while the high-energy emission essentially gets saturated. From the same logic, future instruments observing the source in the TeV range, such as Cherenkov Telescope Array (CTA), are expected to observe even larger source extent and comparatively elevated gamma-ray emission from the HESS J1813-126 source region. 

The results of the B2D model are shown in Fig.~\ref{fig: halo_SED_B2D}. Within the $0.34^\circ$ aperture, the predicted flux reproduces the H.E.S.S. data reasonably well. The required parameters to match the spectrum are an efficiency factor of $\eta \approx 0.22$ and a cutoff energy of $E_{\mathrm{e,c}} \approx 80$~TeV. However, as the aperture is increased beyond the H.E.S.S. extraction region, the integrated SED curves shift upward and begin to overshoot the HAWC and LHAASO measurements. This behavior is a direct consequence of the quasi-ballistic propagation of pairs, which produces flatter SBPs extending to larger radii. In contrast to the sharp drop-off seen in the 2ZISD case, the B2D profiles decline gradually with angle, so that a significant fraction of the flux resides outside the $0.34^\circ$ aperture. As larger integration regions are considered, additional emission from the outer halo is collected. While the effect is most pronounced at lower energies --- where long-lived electrons produce highly extended IC emission --- the flatter SBPs also allow some higher-energy particles to contribute at larger radii, clearly enhancing the high-energy part of the SED as well. Nevertheless, these high-energy electrons are still dominated by radiative losses, so the SBPs continue to retain an energy dependence, i.e., higher-energy profiles remain more compact than those at lower energies. The predicted flux increasingly overshoots the observed HAWC and LHAASO spectra, as well as \textit{Fermi}-LAT upper limits once apertures larger than $0.34^\circ$ are considered. Although the required efficiency is not unphysically larger than unity (as found in some earlier works; see \citealt{wu24}), it is nevertheless high ($\gtrsim 10\%$), which is considerably above typical expectations for pulsar halos. Thus, while the B2D framework can reproduce the H.E.S.S. data, the combination of a high conversion efficiency and the systematic overprediction of the HAWC and LHAASO fluxes indicates that this model provides a poorer overall description of HESS~J1813-126.

\begin{table*}[t]
\centering
\caption{Parameters adopted for the AD model.}
\begin{tabular*}{\textwidth}{@{\extracolsep{\fill}}cccccc}
\hline\hline
$D_{rr, 0}$ (at 100 TeV) [cm$^2$\,s$^{-1}$] & $M_A$ & $\zeta$ ($^\circ$) & $\eta$ & $E_{\mathrm{e,c}}$ (TeV) \\
\hline
$5\times10^{26}$     & 0.20 & 0 & 0.005 & 350 \\
                     & 0.20 & 10 & 0.008 & 200 \\
%\hline
$1\times10^{27}$     & 0.24 & 0 & 0.006 & 300 \\
                     & 0.24 & 10 & 0.010 & 180 \\
%\hline
$5\times10^{27}$     & 0.35 & 0 & 0.011 & 180 \\
                     & 0.35 & 10 & 0.016 & 150 \\
%\hline
$1\times10^{28}$     & 0.43 & 0 & 0.016 & 150 \\
                     & 0.43 & 10 & 0.020 & 130 \\                     
\hline
\end{tabular*}
\label{tab:AD_params}
\end{table*}

Finally, we turn to the results of the AD model. Because the predicted emission is strongly geometry-dependent, we explored the parameter space by considering $M_A = 0.2, 0.24, 0.35$ and $0.43$, and $\zeta = 0^{\circ}$ and $10^{\circ}$. In the AD scenario, the parallel diffusion coefficient $D_{zz}$ must remain consistent with the average Galactic value, so $M_A$ has been chosen in such a way that we can explore different perpendicular diffusion coefficients $D_{rr}$. The two values of $\zeta$ have been chosen to probe different symmetries. As before, assuming an aperture of $0.34^{\circ}$, the SEDs for different combinations of $M_A$ and $\zeta$ are shown in Fig.~\ref{fig: halo_SED_AD_1}, with the corresponding parameter values summarized in Table~\ref{tab:AD_params}. From the table, one can see the efficiency factor $\eta$ (as well as $E_{\rm e,c}$) in the AD model is comparable with those in the 2ZISD model. Additionally, note that for a given $M_A$, increasing $\zeta$ reduces the fraction of projected emission recovered within the chosen aperture and therefore requires even higher efficiency ($\eta$) to match the observed flux. 
For the present work, we restricted ourselves to the representative cases of $M_A = 0.35$, $\zeta = 0^{\circ}$, and $M_A = 0.35$, $\zeta = 10^{\circ}$, and progressively enlarged the aperture (see Fig.~\ref{fig: halo_SED_AD_1}). In the first scenario, the SED rises primarily at lower energies as the aperture is increased, eventually becoming consistent with the observed HAWC, LHAASO, and \textit{Fermi}-LAT fluxes before getting saturated. On the second scenario, although the primary increase in the SEDs remains in the lower energy range, the model SEDs eventually overshoot the HAWC, LHAASO, and \textit{Fermi}-LAT fluxes before saturation, a behavior that can be further understood by examining the model SBPs.
Figure \ref{fig: halo_SED_AD_2} shows the SBPs and gamma-ray morphologies predicted by the AD model. For $\zeta = 0^{\circ}$, where the line of sight is aligned with the ordered magnetic field, the projected gamma-ray emission map is nearly isotropic and circular, and the SBPs extend to large radii with a steep decline, yielding a single, centrally peaked profile in each energy band. In contrast, for $\zeta = 10^{\circ}$, the projected morphology becomes slightly elongated along the magnetic field: the emission extends further in the field direction but falls off more rapidly across it. This broken symmetry introduces distinct profiles, producing multiple curves in the SBPs and a more complex overall structure. 
{From Fig. \ref{fig: halo_SED_AD_2}, it can be seen that the integrated SBP curve corresponding to the azimuthal interval around $\varphi \sim 90^{\circ}$ lies below the others, whereas the same for the azimuthal intervals of $\varphi << 90^{\circ}$ and $\varphi >> 90^{\circ}$ produce comparatively broader, flatter profiles. This happens as the viewing angle increases, the line of sight intersects a longer segment of the field-aligned diffusion region, causing the integrated SBPs in $\varphi << 90^{\circ}$ and $\varphi >> 90^{\circ}$ cases to appear flatter even though the same for $\varphi \sim 90^{\circ}$ case remains steeper.} 
As a result, for $\zeta = 0^{\circ}$ the integrated SED quickly saturates with aperture since most of the flux is already enclosed within the compact, symmetric region, while for $\zeta = 10^{\circ}$ the flatter SBPs allow additional extended emission to contribute at larger radii, leading to a gradual flux increase and a mild overshoot before eventual saturation. The energy dependence further reflects the effect of radiative cooling: in the 8-40 TeV band, the emission is broad and extended, at 40-100 TeV, the profile starts to contract, and at 100-150 TeV, the emission is comparatively more suppressed at large radii, yielding a more centrally peaked profile.

In summary, our exploration of the three transport models underscores both shared features and key distinctions. The 2ZISD scenario provides the most consistent description of the multi-instrument spectra (H.E.S.S., HAWC, LHAASO, \textit{Fermi}-LAT), reproducing the data with reasonable efficiencies and exhibiting an aperture-dependent growth confined primarily to lower energies, without overshooting the observed data. The B2D model can match the H.E.S.S. spectrum within $0.34^{\circ}$, but its flatter profiles lead the integrated SED to exceed the HAWC and LHAASO fluxes at larger apertures, while also requiring comparatively high conversion efficiencies. The AD model naturally captures the geometric imprint of anisotropic diffusion, predicting characteristic SBPs and morphologies testable with future imaging atmospheric Cherenkov telescopes (IACTs). The field-aligned configuration ($\zeta = 0^{\circ}$) explains the H.E.S.S. data well and shows aperture-dependent behavior consistent with multi-instrument observations, whereas the field-misaligned case ($\zeta = 10^{\circ}$) overshoots the fluxes at wider apertures, suggesting that the halo is unlikely to be strongly asymmetric and will likely display spherical morphology. Overall, the 2ZISD framework and the field-aligned AD model -- though grounded in different underlying physics -- both provide a coherent and self-consistent interpretation of HESS J1813-126. Future high-resolution IACT observations will be decisive in distinguishing between these scenarios and constraining particle transport in this system.

A useful way to relate the pulsar central engine PSR J1813-1246 to the surrounding pulsar halo HESS J1813-126 is to compare the high-energy pair budgets implied by the magnetospheric SC emission and by the halo models. In both cases, the model normalizations can be translated into particle injection rates, which we express as dimensionless multiplicities normalized to the Goldreich--Julian (GJ) outflow rate. Note that these multiplicities refer strictly to the particular high-energy subset of the pair population relevant for SC emission and pulsar halo formation in our modeling and they do not represent the global pair yield of the magnetosphere.

The spatial distribution of SC-emitting particles along the gap is given by Eq.~\ref{eq: particle_dist}. Interpreting this as a steady-state occupancy distribution within the emission region, extending from $x_{\rm in}$ to $x_{\rm out}$ with $\Delta \equiv x_{\rm out}-x_{\rm in}$, the mean residence length measured from the inner boundary is
\begin{equation}
x_{\rm res}
\equiv
\frac{\displaystyle \int_{x_{\rm in}}^{x_{\rm out}} (x-x_{\rm in})\,\frac{dN}{dx}\,dx}
{\displaystyle \int_{x_{\rm in}}^{x_{\rm out}} \frac{dN}{dx}\,dx}
=
x_0 - \frac{\Delta}{e^{\Delta/x_0}-1}.
\end{equation}
This expression has the appropriate limiting behaviors: $x_{\rm res}\to x_0$ when the distribution decays steeply ($x_0\ll\Delta$), and $x_{\rm res}\to \Delta/2$ when the distribution is nearly uniform ($x_0\gg \Delta$).

Adopting the standard assumption of ultra-relativistic streaming at $v\simeq c$ along open field lines, the mean residence time is $t_{\rm res} = x_{\rm res}/c$. The associated SC-emitting particle throughput is therefore
\begin{equation}
\dot N_{\rm mag}
=
\frac{N_0}{t_{\rm res}}
=
\frac{c\,N_0}{x_0 - \dfrac{\Delta}{e^{\Delta/x_0}-1}}.
\end{equation}
This quantity represents only the rate of high-energy particles participating in the SC channel and is not intended to describe the full magnetospheric outflow.

For comparison, we normalized $\dot N_{\rm mag}$ to the canonical GJ outflow rate (see, e.g., \citealt{amato24}),
\begin{equation}
\dot N_{\rm GJ}
\simeq
\frac{B_0 R_*^3 \Omega^2}{2 e c},
\end{equation}
with $B_0$ the surface polar magnetic field, $R_*$ the stellar radius, and $\Omega = 2\pi/P$. We therefore defined a high-energy magnetospheric multiplicity:
\begin{equation}
\kappa_{\rm mag}
\equiv \frac{\dot N_{\rm mag}}{\dot N_{\rm GJ}}.
\end{equation}

On the halo side, we adopted an injection power ($\eta\,\dot{E}$) carried by relativistic $e^{\pm}$ with a spectrum given by Eq. \ref{eq: injection_spctrum}. The corresponding particle injection rate is
\begin{equation}
\dot N_{\rm halo}
=
\frac{\eta\,\dot{E}}{\langle E_e\rangle},
\end{equation}
where the mean injected energy is
\begin{equation}
\langle E_e\rangle
=
\frac{\int_{E_{e,\min}}^{E_{e,\max}} E_e\,q_E(E_e)\,dE_e}
{\int_{E_{e,\min}}^{E_{e,\max}} q_E(E_e)\,dE_e}.
\end{equation}
This rate refers only to $e^{\pm}$ above the adopted minimum energy threshold ($E_{\rm e, min}\sim$\,GeV in this case), i.e.,\ the portion of the particles relevant for powering the pulsar halo. Normalizing to the same $\dot N_{\rm GJ}$ gives the halo multiplicity,
\begin{equation}
\kappa_{\rm halo}
\equiv
\frac{\dot N_{\rm halo}}{\dot N_{\rm GJ}}
=
\frac{\eta\,\dot{E}}{\langle E_e\rangle\,\dot N_{\rm GJ}}.
\end{equation}

For our model parameters, {we find $\kappa_{\rm mag} \approx 5.9 \times 10^4$ and} $\kappa_{\rm halo} \approx 2.2 \times10^2$ (corresponding to the 2ZISD case with $D_0 = 5 \times 10^{27} \ \rm cm^2\ s^{-1}$). {The estimated magnetospheric multiplicity is consistent with the typically considered range of $10^3 - 10^5$ in pulsars \citep{timokhin19, amato24, spencer25}.} The halo multiplicity is comfortably below the very large total multiplicities inferred for classical PWNe ($\kappa_{\rm PWN}\sim 10^5$--$10^7$), which arise primarily from low-energy pairs irrelevant for pulsar halo production. Restricted to the high-energy component, the approximate values of $\kappa_{\rm mag}$ and $\kappa_{\rm halo}$ indicate that the magnetosphere does not suffer from a deficit of energetic pairs. This comparison does not, by itself, imply that the SC-emitting region supplies the halo directly; alternative high-energy escape channels such as the equatorial wind or the current sheet may also contribute. Rather, the comparison only shows that the global high-energy pair budget in our model is energetically self-consistent without a need for additional pair-creation channels or exotic accelerators outside the magnetosphere.

\section{Concluding remarks}\label{conclusion}

A compelling case can be made that HESS J1813-126 is indeed a pulsar halo powered by PSR J1813-1246. The multiwavelength phenomenology strongly favors a halo interpretation: no radio or X-ray PWN has been detected despite deep observations \citep{ackermann11, marelli14, guevel25}, which disfavors a confined nebula scenario. The TeV emission is extended ($\sim0.2^\circ$--$0.3^\circ$), with a hard spectrum, and with no shell-like morphology to suggest a SNR or associated MCs \citep{marelli14}. The GeV counterpart detected by \textit{Fermi}-LAT is point-like and coincident with the pulsar, consistent with being dominated by magnetospheric emission. Our phased analysis shows that any residual off-pulse GeV emission is soft and compact, and is more plausibly explained by the imperfect gating of magnetospheric photons rather than an underlying GeV PWN. The extended TeV source HESS J1813-126 has been independently detected by H.E.S.S., HAWC, and LHAASO, all of which confirm its extended morphology and reveal emission up to $>100$~TeV, establishing it as a bright Galactic PeVatron candidate. The positional coincidence with PSR~J1813-1246, an energetic, middle-aged, radio-quiet pulsar located off the Galactic plane, makes a chance alignment highly unlikely. Our transport modeling shows that the observed H.E.S.S. SED can be consistently explained within the pulsar halo framework under multiple plausible transport scenarios (2ZISD, B2D, and AD models).\ Each yields distinct predictions for the SBPs and the aperture-dependent evolution of the SED that provide clear observational diagnostics for future IACTs, yet all converge on the idea that PSR~J1813-1246 has efficiently injected $e^{\pm}$ pairs into the ISM over its lifetime. Taken together, the absence of a classical PWN, the extended and multi-TeV nature of the emission, the spatial coincidence with a powerful pulsar, and the successful reproduction of the data by halo models provide strong evidence that HESS~J1813-126 is best interpreted as a pulsar halo. Recent reanalyses of several extended TeV sources by the H.E.S.S. collaboration have shown that some objects previously reported as moderately extended in the HGPS can exhibit substantially larger and more complex morphologies when studied with improved reconstruction techniques and deeper exposures. Notable examples include HESS J1809-193 and HESS J1813-178, for which updated analyses \citep{aharonian23,aharonian24} revealed broad, multi-degree emission structures. These results have transformed our understanding of evolved PWNe, highlighting scenarios in which a significant fraction of pairs escape the nebula and form very extended pulsar halos, as also discussed for HESS J1809-193 \citep{aharonian23,martin24}. In this broader context, a dedicated, high-statistics H.E.S.S.\ reanalysis of HESS J1813-126 would be particularly valuable, as it may uncover similarly extended emission components and provide decisive constraints on the transport physics operating in this system.

In addition to the halo modeling, we also modeled the high-energy pulsar emission with the SC framework, which successfully reproduces the combined X-ray and gamma-ray spectra of PSR~J1813-1246. The approximate high-energy pair multiplicities found for the magnetospheric and halo populations in our model suggest that the global energetic requirements for both the SC emission and the pulsar halo emission can be met without invoking additional pair-creation channels. This reinforces the association between the pulsar and its surrounding pulsar halo, as both the magnetospheric and halo phenomenology can be explained within a coherent and unified picture.

\begin{acknowledgements}
      A.D.S. thanks the anonymous reviewer for helpful suggestions and constructive criticism, which vastly improved the quality of this work. 
      This work is part of the grant Juan de la Cierva JDC2023-052168-I, funded by MCIU/AEI/10.13039/501100011033 and by the ESF+.
      %                                                                                                                                 %
      This work has been supported by the grant PID2024-155316NB-I00 funded by MICIU /AEI /10.13039/501100011033 / FEDER, UE and CSIC PIE 202350E189. This work was also supported by the Spanish program Unidad de Excelencia María de Maeztu CEX2020-001058-M and also supported by MCIN with funding from European Union NextGeneration EU (PRTR-C17.I1).            
\end{acknowledgements}

   \bibliographystyle{aa} % style aa.bst
   \bibliography{aa.bib} % your references Yourfile.bib

\begin{appendix}
\onecolumn

\section{{\it Fermi}-LAT data analysis}\label{lat}

To investigate the gamma-ray emission from PSR J1813-1246, we analyzed $\sim$16.75 years of P8R3 \texttt{SOURCE} class \textit{Fermi}-LAT data (\texttt{evclass}=128, \texttt{evtype}=3; \citealt{atwood13,bruel18}), covering 2008~August~4 (Mission Elapsed Time (MET) 239557417) to 2025~May~2 (MET 767880664) {in the 1--500~GeV energy range}. {We adopted a minimum energy threshold of 1~GeV to reduce the impact of the bright diffuse emission and to exploit the substantially improved \textit{Fermi}-LAT PSF at higher energies.} The analysis was performed with the \texttt{Fermipy}\footnote{\url{https://fermipy.readthedocs.io/}} package (v1.4.0; \citealt{wood17}), which interfaces with the official \texttt{Fermitools}\footnote{\url{https://fermi.gsfc.nasa.gov/ssc/data/analysis/software/}}. We adopted the instrument response function \texttt{P8R3\_SOURCE\_V3}, excluded events with zenith angle $>105^{\circ}$ to minimize Earth-limb contamination, and used the standard Galactic (\texttt{galdiff}) and isotropic (\texttt{isodiff}) diffuse models, \texttt{gll\_iem\_v07.fits} and \texttt{iso\_P8R3\_SOURCE\_V3\_v1.txt}, respectively. For the source modeling, we adopted the latest \textit{Fermi}-LAT source catalog, 4FGL-DR4 \citep{abdollahi20, ballet23}. 

\begin{figure*}
    \centering
    \includegraphics[width=0.33\textwidth]{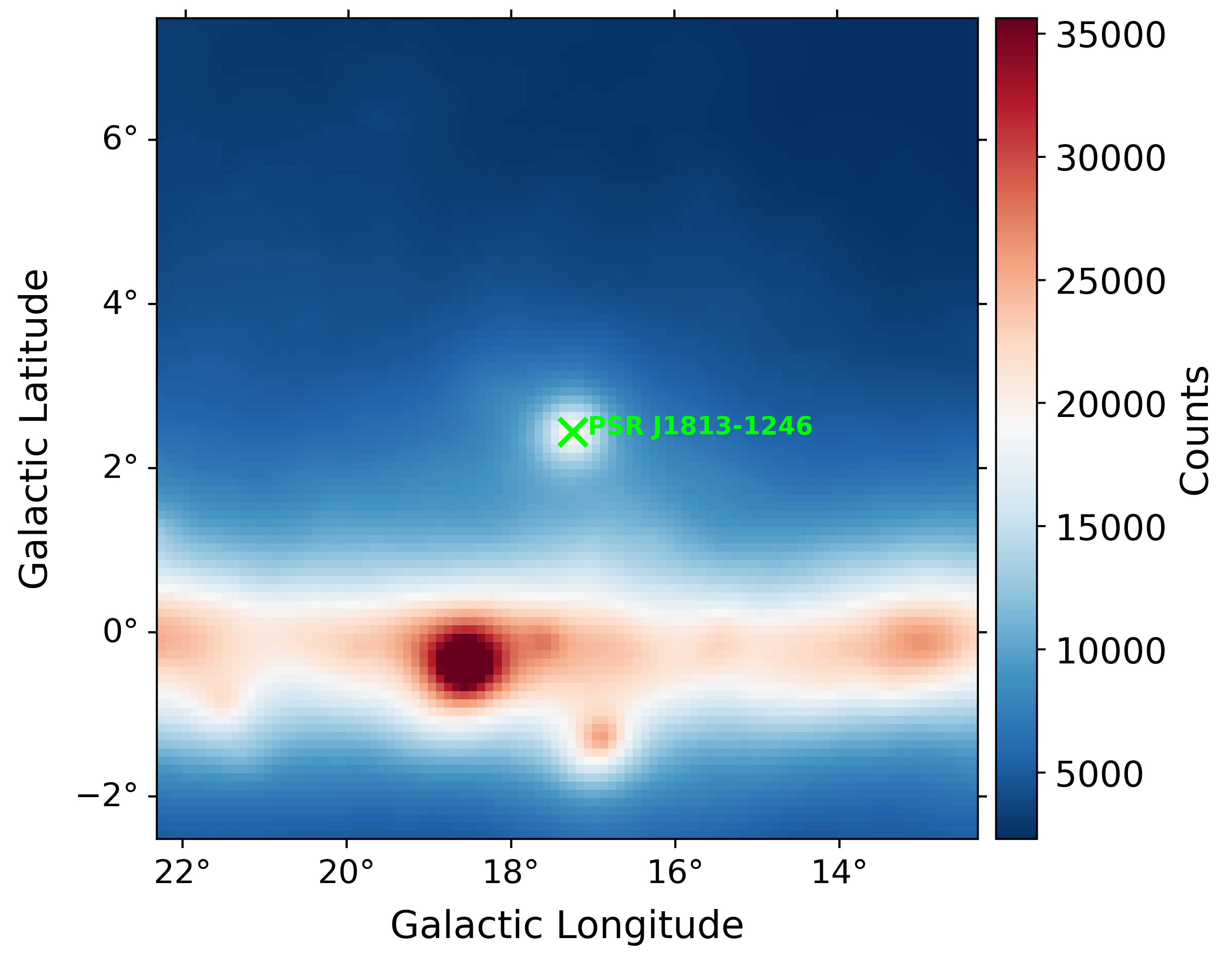}
    \includegraphics[width=0.33\textwidth]{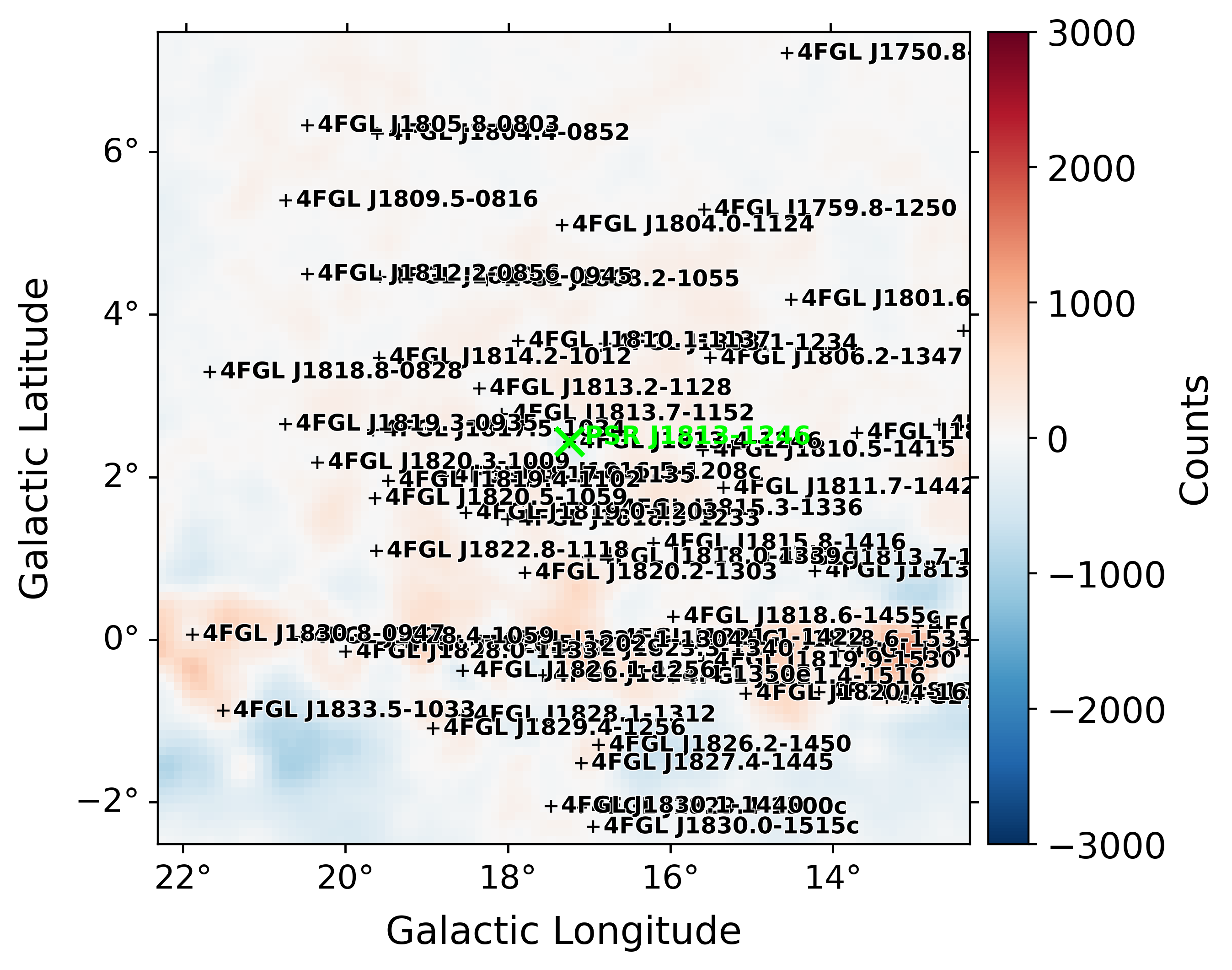}
    \includegraphics[width=0.31\textwidth]{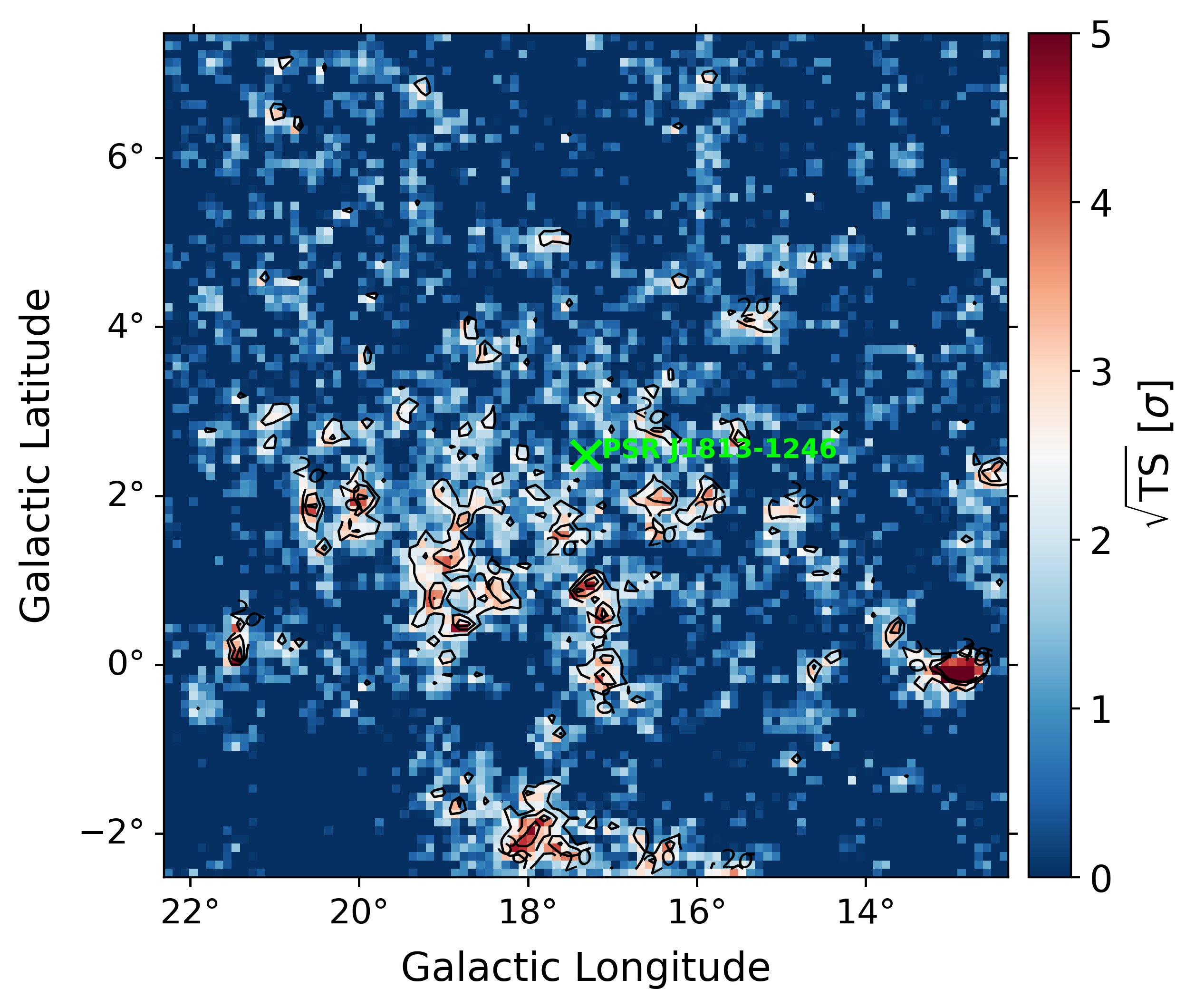}
    \caption{\textit{Fermi}-LAT analysis diagnostic plots. Smoothed data counts map (left panel) and residual (data$-$model) counts map (middle panel) of a $10^{\circ} \times 10^{\circ}$ region centered on PSR J1813-1246 are shown. The residual map includes the positions of all the 4FGL sources in the region (black plus signs). The color bars in both panels indicate the photon counts. {The right panel shows the $\sqrt{TS}$ significance map of the same region, where the color bar indicates the $\sqrt{TS}$ value of the region.} The position of PSR J1813-1246 is marked with a green cross in all panels.}
    \label{fig: counts_map}
\end{figure*}

The region of interest was defined as a circle with $20^{\circ}$ radius centered on PSR J1813-1246, with a spatial bin size of 0.1$^{\circ}$ and 8 energy bins per decade. A binned maximum-likelihood analysis was carried out over a $15^{\circ}\times15^{\circ}$ region centered on the pulsar. We first optimized the baseline model by adjusting the normalizations of diffuse components and 4FGL sources within the analysis region to establish a stable starting point for subsequent likelihood fitting. Then, we freed the normalizations of the 4FGL cataloged sources within $5^{\circ}$ of the pulsar, along with the diffuse components, while fixing distant sources to their 4FGL-DR4 values. To search for un-modeled excesses, we applied \texttt{Fermipy}’s source-finding algorithm, requiring a test statistic (TS) $>16$ and separation of at least $0.2^{\circ}$ from known sources, followed by a global refit by repeating the previous step. From this procedure, the spectral and spatial parameters of all relevant sources were evaluated, and subsequently, the SED, extension, and localization of the target were computed. In Fig. \ref{fig: counts_map}, we show the smoothed data counts map, the residual counts map (with all sources marked), and the $\sqrt{TS}$ significance map of the $10^{\circ}\times10^{\circ}$ region around PSR J1813-1246 to demonstrate that all signal has been properly modeled.

To specifically probe possible GeV PWN emission hidden beneath the pulsar contribution, we performed a pulsar phased analysis using $\sim$11 years of \textit{Fermi}-LAT data, from 2008~August~11 (MET 240162626) to 2019~June~17 (MET 582479577), in {the energy range 1--500 GeV}. We considered $\sim$ 11 years of data in this analysis, corresponding to the time span for which a valid ephemeris was available. Event files with pulse phases were obtained from the \textit{Fermi}-LAT third pulsar catalog (3PC)\footnote{\url{https://fermi.gsfc.nasa.gov/ssc/data/access/lat/3rd_PSR_catalog/3PC_HTML/J1813-1246.html}}, with spacecraft files from the \textit{Fermi}-LAT data server\footnote{\url{https://fermi.gsfc.nasa.gov/cgi-bin/ssc/LAT/LATDataQuery.cgi}}. The analysis was performed in a $15^{\circ}\times15^{\circ}$ region centered on the pulsar position. We followed the same event selection and diffuse models as above. 
The phased analysis proceeded in three main steps. First, we optimized the baseline model as before.
Then, the initial fit was performed by freeing the normalizations of all 4FGL sources within $5^{\circ}$ of the pulsar having a TS $>9$, together with the normalizations of the diffuse templates \texttt{galdiff} and \texttt{isodiff}. Second, we applied the \texttt{Fermipy} source-finding algorithm to search for additional excesses in the analysis region, requiring candidate sources to exceed TS $=16$ and to be separated by at least $0.2^{\circ}$ from cataloged sources to avoid confusion. Third, we refitted the updated source model, performing a global maximum-likelihood optimization to obtain stable convergence.  
During the off-pulse interval, this procedure revealed a new source at the pulsar location, which was then added to the model. Its SED was extracted using the standard \texttt{Fermipy} SED tool, and we tested for possible spatial extension with \texttt{RadialDisk} and \texttt{RadialGaussian} morphologies. We used likelihood ratio tests to refine its localization and to compare point-like and extended-source hypotheses. Note that throughout this work, we have assumed that the extended hypothesis is true only if $TS_{\mathrm{ext}} \geq 25$. The properties of this new source were subsequently examined as potential evidence of faint GeV emission associated with an underlying PWN, emerging once the dominant pulsed magnetospheric emission was gated off.

\section{The synchro-curvature model}\label{pulsar}

For a particle of Lorentz factor $\Gamma$ and pitch angle $\alpha$ moving along a field line of curvature radius $r_{\rm c}$ in a local magnetic field $B$, the gyroradius and SC parameter are
\begin{align}
    r_{\rm gyr} &= \frac{m_e c^2\,\Gamma \sin\alpha}{eB}, \\
    \xi &= \frac{r_{\rm c}}{r_{\rm gyr}}\;\frac{\sin^2\alpha}{\cos^2\alpha}.
\end{align}
The corresponding characteristic photon energy is
\begin{equation}
    E_{\rm c}(\Gamma,r_{\rm c},r_{\rm gyr},\alpha)\;=\;\frac{3}{2}\,\hbar c\,Q_2\,\Gamma^3,
\end{equation}
with
\begin{equation}
    Q_2^2 \;=\; \frac{\cos^4\alpha}{r_{\rm c}^2}\left(1+3\xi+\xi^2+\frac{r_{\rm gyr}}{r_{\rm c}}\right).
\end{equation}
The single-particle spectral power per unit energy at a given position then reads
\begin{equation}
    \frac{dP_{\rm sc}}{dE}
    = \frac{\sqrt{3}\,e^2\,\Gamma\,y}{4\pi\hbar\,r_{\rm eff}}
      \left[(1+z)F(y) - (1-z)K_{2/3}(y)\right],
\end{equation}
where
\begin{equation}
    y\equiv \frac{E}{E_{\rm c}},\qquad
    z = (Q_2 r_{\rm eff})^{-2},\qquad
    r_{\rm eff}=\frac{r_{\rm c}}{\cos^2\alpha}\left(1+\xi+\frac{r_{\rm gyr}}{r_{\rm c}}\right)^{-1},
\end{equation}
and
\begin{equation}
    F(y) \;=\; \int_y^\infty K_{5/3}(y')\,dy' ,
\end{equation}
with $K_\mathrm{n}$ being the modified Bessel functions of the second kind of index n. The total SC power radiated by a single particle is
\begin{equation}
    P_{\rm sc} \;=\; \frac{2 e^2 c \Gamma^4}{3r_{\rm c}^2}\;g_r, 
    \qquad
    \mathrm{where}
    \qquad
    g_r \;=\; \frac{r_{\rm c}^2}{r_{\rm eff}^2}\;
               \frac{1+7(Q_2 r_{\rm eff})^{-2}}{8(Q_2 r_{\rm eff})^{-1}}.
\end{equation}

The evolution of $\Gamma$, $\alpha$, and $\xi$ is obtained by solving the coupled differential equations\begin{align}
    \frac{d(p\sin\alpha)}{dt} &= -\,\frac{P_{\rm sc}}{c}\,\sin\alpha, \\
    \frac{d(p\cos\alpha)}{dt} &= \,eE_\parallel - \frac{P_{\rm sc}}{c}\,\cos\alpha,
\end{align}
with $p (=\Gamma m_e c)$ being the relativistic momentum and $x$ ($dx=c\,dt$) being the gap coordinate. We adopted $\Gamma_{\rm in}=10^3$ and $\alpha_{\rm in}=45^\circ$ as initial conditions. The average SC spectrum throughout the trajectory is then
\begin{equation}
    \frac{dP_{\rm tot}}{dE} = \int_{x_{\rm in}}^{x_{\rm out}}\frac{dP_{\rm sc}}{dE}(x)\;
        \frac{dN}{dx}(x)\;dx,
\end{equation}
with an effective particle distribution
\begin{equation}\label{eq: particle_dist}
    \frac{dN}{dx}(x)\;=\;\frac{N_0\,e^{-(x-x_{\rm in})/x_0}}{x_0\left(1-e^{-(x_{\rm out}-x_{\rm in})/x_0}\right)}.
\end{equation}
In our setup, the inner boundary ($x_{\rm in}$) is at $0.5\,R_{\rm lc}$ and the outer boundary ($x_{\rm out}$) is at $1.5\,R_{\rm lc}$ of the emission region, and $x_0$ is the length scale, which was considered a free parameter. The magnetic field and curvature radius along the path are parameterized as
\begin{equation}
    B(x) = B_0 \left(\frac{R_*}{x}\right)^{b_{sc}},\qquad
    r_{\rm c}(x) = R_{\rm lc}\left(\frac{x}{R_{\rm lc}}\right)^{\eta_{sc}},
\end{equation}
where $R_*$ ($\approx$ 10 km) is the pulsar radius, $B_0 (\approx 6.4 \times10^{19} \sqrt{P\,\dot P} \ \mathrm{G})$ is the dipolar magnetic field at the polar surface, $R_{\rm lc}$ ($ = Pc/2 \pi$) is the light cylinder radius and $\eta_{sc}=0.5$. The magnetic field gradient $b_{sc}$, which determines how quickly the magnetic field decreases along particle trajectories, along with the $N_0 (=\int^{x_{\mathrm{max}}}_{x_{\mathrm{min}}}\frac{dN}{dx} dx)$, denoting the total number of particles and used to set the normalization of the spectrum, were also kept free.

\section{The pulsar halo particle transport models}\label{halo}

The evolution of the lepton distribution $N(E_e, r,t)$ is governed by the time-dependent, diffusion-loss equation
\begin{equation}
\frac{\partial N}{\partial t} \;=\; \nabla\!\cdot\!\left(D(E_e, r)\,\nabla N\right)\;+\;\frac{\partial}{\partial E_e}\!\left(b(E_e)\,N\right)\;+\;Q(E_e, r,t),
\label{eq:transport}
\end{equation}
where $D$ is the diffusion coefficient, $b(E_e)=|dE_e/dt|$ is the absolute energy-loss rate, and $Q$ is the source term. For relativistic leptons above GeV energies, the losses are quadratic in energy, $b(E_e)\approx b_0 E_e^2$, with $b_0$ determined by the sum of synchrotron and IC contributions. The IC term is evaluated for each photon field with the full Klein–Nishina cross section.

The source term $Q$ can be divided into spatial ($q_r$), temporal ($q_t$), and energy components ($q_E$). The spatial component of injection is taken to be point-like in space at the pulsar position, i.e., $q_r(r)=\delta(r)$. The time dependence follows the pulsar spin-down power in the dipole case ($n=3$):
\begin{equation}
q_t(t) = \frac{(1+t/\tau_0)^{-2}}{(1+t_s/\tau_0)^{-2}}, \qquad t \geq 0,
\end{equation}
where $\tau_0$ is the initial spin-down timescale and $t_s$ the source age. The spectrum of injected particles at the current epoch is assumed as
\begin{equation}\label{eq: injection_spctrum}
q_E(E_e)=q_{E,0}\,E_e^{-p}\,\exp\!\left[-\left(\frac{E_e}{E_{e, c}}\right)^{s}\right],
\end{equation}
within $E_{\mathrm{e, min}} \approx 1 \ \mathrm{GeV}$ and $E_{\mathrm{e, max}} \approx 1 \ \mathrm{PeV}$ with normalization set by an efficiency factor $\eta$ through $\eta\,\dot E=\int q_E(E_e)\,E_e\,dE_e$.

Once $N(E_e,r)$ is obtained by solving Eq. \ref{eq:transport}, the observable gamma-ray signal is calculated. For the spherically symmetric case, the surface distribution is calculated by performing line-of-sight (LOS) integration on the electron number density:
\begin{equation}\label{eq_IC_1}
S_e(E_e,\theta)=\int_{0}^{\infty} N(E_e,r)\,dl,
\end{equation}
with $r=\sqrt{r_s^2+l^2-2r_s l \cos\theta}$, where $r_s$ is the distance to the pulsar, $l$ is the distance from a point to the observer along LOS, and $\theta$ is the angle between LOS and the direction from the observer to the pulsar. The IC surface brightness is then
\begin{equation}\label{eq_IC_2}
s_\gamma(E_\gamma,\theta) = \frac{1}{4\pi}\int dE_e \int d\epsilon \;
S_e(E_e,\theta)\,f(E_e,\epsilon,E_\gamma),
\end{equation}
where $f$ signifies the production rate of emitted gamma-ray photons with energy $E_{\gamma}$ from the interaction between electrons of energy $E_e$ and target photons of energy $\epsilon$. 
The energy spectrum within an angular range can be calculated using
\begin{equation}\label{eq: theta_integrated_spectrum}
    F(E_{\gamma}) = \int_{\theta_{\gamma,1}}^{\theta_{\gamma,2}} s_\gamma(E_\gamma,\theta)\ 2 \pi\theta\ d\theta
,\end{equation}
whereas the energy band-integrated SBP is
\begin{equation}
S_\gamma(\theta)=\int_{E_{\gamma,1}}^{E_{\gamma,2}} s_\gamma(E_\gamma,\theta)\,w(E_\gamma)\,dE_\gamma,
\end{equation}
with $w(E_\gamma)=1$ for photon counts or $w(E_\gamma)=E_\gamma$ for energy flux. Finally, this intrinsic profile must be convolved with the instrument’s PSF:
\begin{equation}\label{eq:PSF}
\tilde S_\gamma(\theta) = \int_0^\infty \int_0^{2\pi}
S_\gamma(\theta') \,
g\!\left(\theta, \theta', \phi\right)\,
\theta' \, d\phi \, d\theta',
\end{equation}
where $(\theta,\theta',\phi)$ denote polar coordinates in the plane of the sky, with $\theta$
and $\theta'$ being the observed and intrinsic angular distances from the pulsar, respectively,
and $\phi$ being the azimuthal angle. $g$ is the PSF. In this work, a symmetric 2D Gaussian PSF with width $\sigma_{PSF}$ is considered. For a cylindrical model, such as the AD model, the LOS integration takes the form
\begin{equation}
S_e(E_e,\phi,\theta)=\int_{0}^{\infty} N(E_e,r, z)\,dl,
\end{equation}
where the mapping of $(l, \theta, \phi) \rightarrow (r, z)$ has been done following \cite{liu19}. We briefly discuss the underlying idea and corresponding equations of the three models below.

\subsection{Two-zone isotropic suppressed diffusion model}
In isotropic cases, the diffusion-loss equation reduces to spherical symmetry:  
\begin{equation}
\frac{\partial N (E_e, r, t)}{\partial t} =
\frac{1}{r^2}\frac{\partial}{\partial r}
\!\left(r^2 D(E_e,r)\frac{\partial N}{\partial r}\right)
+ \frac{\partial}{\partial E_e}\!\left(b(E_e)N\right) + Q(E_e,t)\delta(r).
\label{eq:transport_sph}
\end{equation}
A single-zone suppressed diffusion model is generally inconsistent with the B/C ratio and cosmic-ray electron data \citep{aguilar16, hess17, hooper18}, motivating the two-zone approach \citep{hooper17, fang18, profumo18, tang19, johannesson19}. This introduces a slow-diffusion bubble of radius $r_{2{\rm z}}$ (tens of parsecs) around the pulsar,
\begin{equation}
D(E_e,r)=
\begin{cases}
D_{\rm in}(E_e)= \Xi_{2\mathrm{z}}\,D_{\rm ISM, 0} \left(\tfrac{E_e}{100~\mathrm{TeV}}\right)^{\delta}, & r \leq r_{2{\rm z}}, \\
D_{\rm out}(E_e)= D_{\rm ISM} = D_{\rm ISM, 0} \left(\tfrac{E_e}{100~\mathrm{TeV}}\right)^{\delta}, & r > r_{2{\rm z}},
\end{cases}
\label{eq:2Z_diff}
\end{equation}
with $\Xi_{2\mathrm{z}} \ll 1$. In this work, we adopted $r_{2{\rm z}} = 50$ pc \citep{xi19, wu24}. The diffusion coefficient $D_{\rm ISM, 0} = 3.16 \times 10^{29} \ \mathrm{cm^2 \ s^{-1}}$ at 100 TeV, and Kolmogorov-like $\delta = 0.3$ were used unless otherwise stated. This configuration reproduces the compact halos observed by HAWC around Geminga and Monogem \citep{abey17, hooper17}. Unlike the B2D or AD models, it enforces a sharp radial break in $D(E_e,r)$ with suppressed diffusion within $r_{2{\rm z}}$. This suppressed diffusion ensures particles cool before escaping, yielding a bright inner core and a drop in brightness beyond $r_{2{\rm z}}$. Cosmic-ray self-generated turbulence, via resonant and nonresonant streaming instabilities, has been proposed as the origin of this slow-diffusion region \citep{skilling71, skilling75, bykov13, evoli18a, evoli18b, mukho22, fang19, bell04, gupta21, schroer21}. Although semi-analytical solutions are present for the 2ZISD model \citep{osipov20}, we used the finite-volume discretization method \citep{fang18} to solve Eq. \ref{eq:transport_sph} with discontinuous diffusion coefficients in Eq. \ref{eq:2Z_diff}.

\subsection{Ballistic-to-diffusion model}
The one-zone normal diffusion model with spherical transport (Eq.~\ref{eq:transport_sph}) suffers from an unphysical superluminal problem. With mean free path $\lambda(E_e) = 3D(E_e)/c$, freshly injected particles cannot diffuse for $t \leq \lambda/c$; otherwise, their average displacement, $\sqrt{4Dt}$, would exceed that at light speed \citep{aloisio09}. A more physical picture is ballistic propagation at $r \ll \lambda$ and diffusion at $r \gg \lambda$, implying a transition between regimes.  

This can be treated by introducing the J\"{u}ttner propagator \citep{aloisio09}, which preserves causality and interpolates between ballistic and diffusive transport. The diffusion-loss solution then reads
\begin{align}
N(E_e,r) &= \int dt_0 \,\frac{q_t(t_0)\,q_E(E_{e,\star})}{b(E_e)} \,
\frac{b(E_{e, \star})}{4\pi [c(t-t_0)]^3} \nonumber\\
&\times \frac{H(1-\xi_1)\,\kappa_1}{(1-\xi_1^2)^2\,K_1(\kappa_1)}\,
\exp\!\left[-\frac{\kappa_1}{\sqrt{1-\xi_1^2}}\right],
\end{align}
where $\xi_1=r/[c(t-t_0)]$, $\kappa_1=[c(t-t_0)]^2/(2\lambda_\star)$, $\lambda_\star = \int_{E_e}^{E_{e, \star}}D(E'_e)\,dE'_e/b(E'_e)$, and $K_1$ is the first-order modified Bessel function of the second kind.  

During quasi-ballistic propagation, the velocity distribution is anisotropic, so IC emission assuming isotropy (Eq. \ref{eq_IC_1}) is invalid. A correction is included in the LOS integration:
\begin{equation}
    S_e(E_e, \theta) = \int^{\infty}_0 2\, N(E_e, r)\, M(x, \mu)\, dl,
\end{equation}
with \citep{aharonian10, prosekin15}
\begin{equation}
    M(x, \mu) = \frac{1}{Z(x)} \exp\!\left[-\frac{3(1-\mu)}{x}\right],
\end{equation}
where $x = rc/D(E_e)$, $Z(x) = \tfrac{x}{3}[1 - \exp(-6/x)]$, and $\mu = (r_s\cos \theta -l)/r$. The IC emission was then evaluated with Eq.~\ref{eq_IC_2}.  

\cite{recchia21} demonstrated that the B2D framework can reproduce the multi-TeV SBPs of pulsar halos using a standard ISM diffusion coefficient, albeit at the cost of invoking larger pair conversion efficiencies compared to the 2ZISD model. In this scenario, the halo can appear more extended, with a possible shrinkage at high energies due to the limited ballistic propagation length and radiative cooling effects \citep{prosekin15, yang22}. In the B2D model, the apparent halo size is mainly set by the ballistic propagation length of pairs before they isotropize and enter the diffusive regime, whereas in the 2ZISD model it is controlled by the combined effects of radiative cooling and the diffusion distance of the pairs.

\subsection{Anisotropic diffusion model}
Given that interstellar turbulence is sub-Alfv\'{e}nic (Alfv\'enic Mach number $M_A < 1$), and the magnetic field maintains a mean direction within one coherence length of 10 - 100 pc, transport is no longer isotropic within that length, since particle diffusion along the mean magnetic field direction is much faster than that in the perpendicular direction. In this scenario, the perpendicular diffusion coefficient $D_{rr}$ is suppressed by a factor $M_A^4$ with respect to the parallel diffusion coefficient $D_{zz}$ \citep{yan08}, assuming the mean magnetic field is oriented along the $z$ axis. For this model, the diffusion–loss equation then takes the cylindrical form
\begin{equation}
\begin{split}
\frac{\partial N (E_e, r, z, t)}{\partial t} &=
\frac{1}{r}\frac{\partial}{\partial r}\!\left(r D_{rr}(E_e)\frac{\partial N}{\partial r}\right) + D_{zz}(E_e)\frac{\partial^2 N}{\partial z^2} \\
&\quad 
+ \frac{\partial}{\partial E_e}\!\left(b(E_e)N\right) + Q(E_e, t)\,\delta(r)\,\delta(z),
&
\end{split}
\label{eq:transport_cyl}
\end{equation}

with
\begin{equation}
D_{rr}(E_e)=D_{rr, 0} \left(\frac{E_e}{100~\mathrm{TeV}}\right)^{\delta}, \qquad
D_{zz}(E_e)=M_A^{-4}\,D_{rr}(E_e).
\end{equation}
After rescaling $z' = M_A^2 z$, the diffusion becomes isotropic in $(r,z')$ coordinates, giving
\begin{equation}
N(E_e,r,z)=\int dt_0 \,\frac{q_t(t_0)\,q_E(E_{e, \star})}{b(E_e)} \,
\frac{b(E_{e, \star})\,M_A^2}{(4\pi \lambda_\star)^{3/2}}
\exp\!\left[-\frac{r^2+M_A^4 z^2}{4\lambda_\star}\right],
\end{equation}
where $\lambda_\star = \int_{E_e}^{E_{e, \star}}D_{rr}(E'_e)\,dE'_e/b(E'_e)$.
The AD model predicts a distinctive observational signature of elongated, asymmetric halo morphology that depends on the viewing angle $\zeta$ relative to the ordered field. \citet{liu19} demonstrated that the multi-TeV SBP of Geminga can be reproduced if $\zeta \leq 5^{\circ}$ with respect to the mean magnetic field. The AD model removes the need for suppression via turbulence amplification as present in the 2ZISD model. AD and 2ZISD models typically yield comparable conversion efficiencies, as both produce steep SBP declining at small radii due to reduced particle diffusion, though the physical origin of slow diffusion differs. Nevertheless, it is to be noted that the consistency of the AD model in light of the general pulsar halo observations is still debatable \citep{dela22}.

\end{appendix}

\end{document}